\documentclass[twocolumn,secnumarabic,amssymb,preprint nobibnotes, aps, prb]{revtex4-2}
\pdfoutput=1
\usepackage{graphicx}
\usepackage{amsmath}

\begin{document}


\title{ RKKY interaction on curved surfaces: Different behavior of Dirac and Schr\"{o}dinger carriers as interaction mediating particles }


\author{Behzad Shokri}

\affiliation{%
	Department of Physics, Azarbaijan Shahid Madani
	University, 53714-161, Tabriz, Iran
}
\author{Ali Eghbali}%

\affiliation{%
	Department of Physics, Azarbaijan Shahid Madani
	University, 53714-161, Tabriz, Iran
}%


\author{Arash Phirouznia}
\email[Corresponding author's Email: ]{Phirouznia@azaruniv.ac.ir}
\affiliation{%
	Department of Physics, Azarbaijan Shahid Madani
	University, 53714-161, Tabriz, Iran 
}%
\altaffiliation{Condensed Matter
	Computational Research Lab. Azarbaijan Shahid Madani University,
	53714-161, Tabriz, Iran}

\date{\today}

\begin{abstract}
The RKKY interaction between two impurities on a curved surface is investigated. Practical curved sample parameters are
 considered for investigating curvatures, similar to those observed experimentally in available samples such as graphene.
For this type of two-dimensional curved systems and for two types of  Schr\"{o}dinger and Dirac carriers, the RKKY interaction and local density of states (LDOS)
are calculated. Results show that the strength of the RKKY interaction on a curved surface depends significantly on the nature of particles embedded on this surface, indicating that the
geometry has a different effect on the dynamics of Schr\"{o}dinger and Dirac carriers. While the curved surface increases the RKKY interaction strength for Schr\"{o}dinger-type carriers in comparison with the flat sample, RKKY interaction decreases in the curved surface for Dirac-type particles. This can be understood regrading the effective magnetic field experienced by the Dirac particles as a result of the curved surface. There is also a
correlation between LDOS and the RKKY interaction, which could be explained by the effect of conducting
electrons density on the exchange interaction mediated by these carriers.
\end{abstract}


\maketitle


\section{Introduction}
Following the discovery of graphene and its amazing properties, a rich field of studies on two-dimensional (2D) surfaces embedded in three-dimensional (3D) spaces was opened up. For decades, condensed matter physicists have been interested in the dynamics of carriers restricted to 2D materials. Freestanding graphene has piqued the interest of a broader audience due to the enormous potential for its applications in fields such as photonics, spintronics, and sensor technology \cite{6}. Experimental observations, such as transmission electron microscopy  and scanning tunneling microscopy , have revealed that samples of freestanding graphene have random spontaneous curves that can be considered ripples of varying sizes, several angstroms in height and several nanometers in length \cite{1,2}. Therefore theoretical results could be corrected by taking these ripples into account.
\\ 
Quantum theory has been applied on a small scale in Euclidean space without curvature, as well as in the study of black holes, and also on a large scale in curved spaces. The discovery of graphene, provides a practical way of understanding quantum theory on small-scale curves. Meanwhile,  it is difficult to generalize quantum mechanics formalism to curved spaces because each of the present approaches has its own flaws \cite{13}.
\\
Nanostructures have established an experimental field that may provide direct evidence for the realization of geometry-induced quantum effects. The fabrication of nanoscale and microscale surfaces opens up ways for the quantum theory of particles confined to curved structures \cite{13}.
\\
The conventional method of the da Costa confining potential, known as thin-layer quantization \cite{12}, was used to insert the constraint of sample to provide a specific 2D curved system. This method is consistent with the uncertainty principle and avoids the ambiguous problem of arranging operators. There are two approaches for studying the motion of a particle confined to the space of a manifold embedded in Euclidean space \cite{6}. The classical Hamiltonian is given in terms of generalized momentum in the conventional quantization approach, which considers the motion of a particle in a preliminary manifold, and the system is quantized using the canonical method. The motion of the particle in this method is solely determined by the geometry of the manifold and is independent of the space in which the manifold is embedded. In the confining potential approach, on the other hand, the particle is confined by a strong force acting perpendicular to the manifold. By reducing the motion perpendicular to the surface, an effective Hamiltonian for particle motion on the surface is obtained. The geometry of the manifold boundary surface determines the effective Hamiltonian. The ambiguity of the quantum operators ordering perplexes the canonical quantization, allowing for several compatible quantum approaches that differ just in a term proportional to the scalar curvature of the manifold boundary \cite{6}. However, the effective confining approach depends on the physical mechanism of the confining process. As a result, the strong confining potential model is more compatible with real physical systems.
\\

One common issue in the field of spin-transferred magnetic order is how magnetic moments of impurities embedded in nanoscale systems could interact with each other even when they are far apart. This interaction is the dominant exchange magnetic interaction in metals where the overlap between neighboring electrons is small or not direct. In this case impurities interact with each other through an intermediary which in metals are the conduction electrons. This type of exchange was first proposed by Ruderman and  Kittel \cite{3} and later extended by Kasuya and Yossida, and this theory is now commonly known as the RKKY interaction. In this paper, the effect of geometry on quantum aspects is investigated by studying the influence of curvature on  RKKY interaction of free-standing graphene sheets with Dirac carriers. Here, we also examine the RKKY interaction in a 2D curved surface embedded in the 3D space with Schr\"{o}dinger carriers. This can reveal some features of the relationship between geometry and quantum mechanics.
\\
For this purpose, two different types of 2D samples have been considered. We use the same model that has been employed for curved graphene \cite{5}, in which carriers are Dirac quasi-particles and also a thin film of curved surfaces, in which carriers follow the Schr\"{o}dinger equation. It has been assumed that the ripple is a Gaussian bump located in the center of the flat surface \cite{5}. 
\\
The influence of the smooth curves, specified in the experiments, on the RKKY interaction is investigated here. The dynamical equations of the carriers in the flat surface state must be rectified using methods for a curved surface state. Two approaches have been proposed for the analytical study of real corrugated graphene samples. The first approach is a combination of tight-binding and elasticity theory \cite{31new} and the second is a formulation of quantum field theory in curved space \cite{5,6}. The presence of ripples observed experimentally by TEM, led us to use a model to correct calculations performed without considering these ripples. 
\\
In this paper, regarding the modified dynamics of confined carriers, the influence of the roughness on the RKKY interaction is obtained. This approach is used in both cases mentioned above, i.e., ordinary curved surfaces of small thickness in which the Schr\"{o}dinger equation must be modified and the other one is the curved graphene surface, in which the Dirac equation must be modified. In the non-relativistic case, the Costa formalism is applied to correct the Schr\"{o}dinger equation of carriers confined in a curved surface. This approach is called the thin layer-quantization (a confining potential formalism) \cite{6,8,12,15,32}. In the case of curved graphene surfaces, no effective geometric potential is required to confine the charge carriers. Therefore, curvature manifests itself in the  Dirac matrices and the partial derivatives have been converted to the covariant derivative. This kind of effect arises from the spinor nature of the wave function of graphene that comes from the two-sublattice nature of graphene structure, where, it is assumed that the dynamics of carriers in 2D curved space such as graphene in low energies follows the Dirac equation in (2 + 1)-dimensional curved space-time \cite{5}.
\\ 
In the following, we obtain the propagator in both cases, i.e. for Dirac and Schr\"{o}dinger particles by using the conventional perturbation approach, we can obtain the RKKY interaction. Then the results of the RKKY interaction in the curved surface are compared with the results of the flat surface.
\\
RKKY interactions of flat graphene are presented in several studies \cite{9,10,11}.  
It has been shown that the intensity of the interaction disappears faster than the other 2D materials \cite{9}. The interaction decays as $\frac{1}{d^3}$  or faster in undoped graphene, where $d$ is the separation distance between the magnetic elements. In the case of doped graphene, the interaction decays with the asymptotic relationship of  $\frac{1}{d^2}$   as expected for 2D materials, but it has not yet been seen experimentally \cite{9}. One can conclude that the RKKY interaction is short range. For example, the intensity of RKKY interaction in carbon nanotubes is predicted to decay as $\frac{1}{d}$  \cite{9}. Flat doped graphene belonging to the same subunits is ferromagnetic and otherwise antiferromagnetic \cite{11}. 
\\
This paper is organized as follows: In Sec. II, we look at the modeling of bumps and approach. In Sec. III, the thin layer-quantization approach is employed to get the modified Schr\"{o}dinger equation for curved surfaces. In Sec. IV, the modified Dirac equation for curved surfaces is obtained. The results and discussion of the RKKY interaction computations in both cases is represented in Sec. V. Finally, the conclusions is given in Sec. VI.
\section{MODEL AND APPROACH}
While
 the actual origin of ripples is unknown, their shape has been observed experimentally\cite{1,2}. In Refs. \cite{5,15}, a general metric was used for studying the case of smooth curvatures that have no singularity throughout the surface and becomes asymptotically flat. We look at the general scenario of a smooth protuberance that fits without singularities in the flat surface's center and is relatively compatible with the free-standing graphene sheets mentioned \cite{1,5}. Consider, for example, a surface immersed in 3D space with polar symmetry described in the cylindrical coordinate system. Polar coordinates of surface projection onto the $z=0$ plane are used to create this surface, which is defined as follows: 
\begin{eqnarray}
	z(r) = Ae^{-(r/b)^2}.
	\label{eq1}	
\end{eqnarray}
Here, $z(r)$  represents the height of the curved surface as compared with the flat surface $z(r) = 0$, \cite{5}. The parameter $A$ stands for the height and $b$ represents the mean width of the Gaussian surface. By considering the position vector in a cylindrical coordinate system in the form,
\begin{eqnarray}
	\vec{r}=\hat{r}r+\hat{k}z,
	\label{eq2}	
\end{eqnarray}
and by using the fact that
\[g_{\mu\nu}=-\frac{d\vec{r}}{d{{q}^{\mu}}}.\frac{d\vec{r}}{d{{q}^{\nu}}},\]

then the metric can be obtained as follows:     
\\	
\begin{eqnarray}
	g_{{\mu}{\nu}} =-\begin{pmatrix}
		1+4{{\left(\frac{A}{b} \right)}^{2}}{{\left(\frac{r}{b} \right)}^{2}}{{e}^{-2{{\left(\frac{r}{b} \right)}^{2}}}} & 0  \\
		0 & {{r}^{2}}  \\
\end{pmatrix},
\label{eq3}
\end{eqnarray}
\begin{figure*}[htbp]
	
	\centering
	
	\includegraphics[width=19cm,height=8cm]{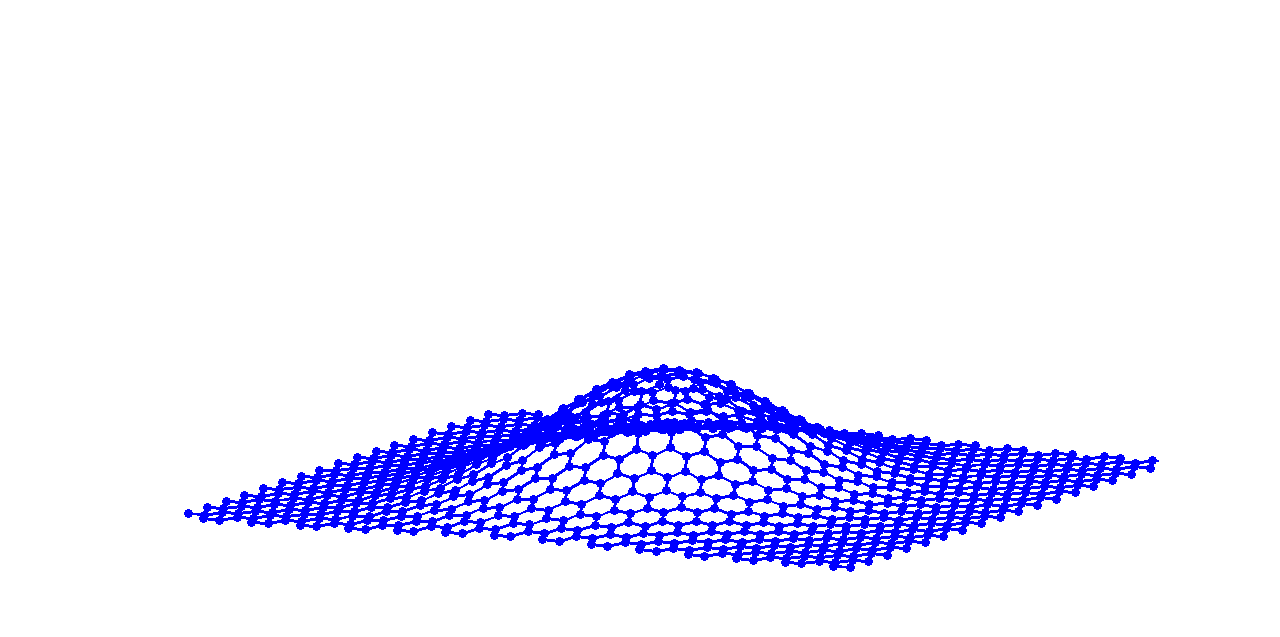}
	\caption{\texttt (Color online) The smooth curved graphene with a Gaussian bump. }
	\label{fig:1}

\end{figure*}
in which ${{g}_{rr}}$ depends on $( \frac{A}{b})^{2}$, square ratio of the height to the mean width of the Gaussian surface, that shows the topographic characteristics of the surfaces. We have introduced $f(r)=4{{\left( r/b \right)}^{2}}{{e}^{-2{{\left(r/b \right)}^{2}}}} $ and             
$~\alpha ={{\left(A/b\right)}^{2}}.$   Experimental values reported for $b$ (width) is in a range from $2$ nm to $20$ nm, while $A$ (height) could be from $0.1$ nm up to $0.5$ nm \cite{1,2}. Figure \ref{fig:1} depicts a typical smooth graphene Gaussian bump.
\\

According to Minkowski space, the spatial  component of displacement is
\begin{eqnarray}
	\label{metric}
d{{s}^{2}}&=&d\vec{r}.d{{\vec{r}}^{*}}\nonumber\\&=&
d{{q}^{\mu}}d{{q}_{\mu}}\\&=&
{{g}_{\mu\nu}}d{{q}^{\mu}}d{{q}^{\nu}}\ 
	\nonumber\\&
=&-(1+\alpha f(r))d{{r}^{2}}-{{r}^{2}}d{{\theta }^{2}}\nonumber
\label{eq4}.
\end{eqnarray} 
The Green’s function of the Schr\"{o}dinger and Dirac carriers is used to compute the LDOS and the RKKY interaction. The following equations are employed to obtain the above elements using the Green’s function. The influence of the curved surface on RKKY interaction could then be captured by the Green's functions \cite{10}. For a perturbation, 
$ V=-\lambda \vec{S}.\vec{s}\delta (\vec{r})$, where $\vec{S}$ is the spin of the impurity and $\vec{s}$ is the spin of the conduction electron, the energy is given by
 \begin{eqnarray} E(R)&=&J{{\vec{S}}_{1}}.{{\vec{S}}_{2}}\nonumber\\&=&\frac{{{\lambda }^{2}}{{\hbar }^{2}}}{4}\chi (r,{r}'){{\vec{S}}_{1}}.{{\vec{S}}_{2}},
 	\label{eq5}
 \end{eqnarray}
where $\vec{S}_{1}$, $\vec{S}_{2}$ are the spins of impurities. The unperturbed retarded Green's functions are used to express the standard susceptibility $\chi (r,{r}')$ :\\
\begin{eqnarray}
\chi (r,{r}')=\frac{-2}{\pi }\int_{-\infty }^{{{E}_{f}}}{dEIm[G(r,{r}',E)G({r}',r,E)]}.    \label{eq6} 
\end{eqnarray}                
We can obtain by continuing
{\small \begin{eqnarray}
	{J}_{_{RKKY}}(r,r')=-\frac{\lambda ^2}{2\pi }{\hbar^2}\int_{-\infty }^{E_f}{dE~Im\Big[G(r,r',E)G(r',r,E)\Big]}.~~~~
	\label{eq7}
	\end{eqnarray} }           
The local density of states can be obtained as follows:
\\
\begin{eqnarray}
\rho(\vec{r},E)=-\frac{1}{\pi }ImTr\Big[G(\vec{r},\vec{r},E)\Big].  
\label{eq8} 
\end{eqnarray}          
\section{Modified Schr\"{o}dinger equation of Fermions on curved surface }
The problem is not complicated in the absence of curvature, i.e. in the case of the Euclidean geometry where the solutions of the Schr\"{o}dinger equation are plane waves, which are eigenfunctions of the linear momentum operator. Furthermore, these plane waves are the eigenfunctions of momentum and energy operators at the same time. If the curvature of space is continuous and non-zero, the canonical and Noether momenta do not match \cite{Bracken2,27}, so the Noether momenta have no Poisson bracket, as expected for the quantum versions of these quantities, the operators. In every space with a variable or non-zero curvature, these characteristics complicate the problem. The plane wave is a Euclidean concept and it is unclear how to apply it to curved space \cite{27}.\\
Two modifications must be applied to Schr\"{o}dinger's equation to describe the quantum mechanics of a particle confined to a curve or a surface in ${{\mathbb{R}}^{3}}$. The surface metric requires that the kinetic energy operator should be adjusted. The potential energy term have to be modified by the addition of a geometric potential, which is required to limit the particle to the curve or surface.
The total potential $V={{V}_{physical}}+{{V}_{geometric}}$ is the sum of two terms that make up potential energy.
External fields (electrical and magnetic) provide the physical potential and geometrical potential is created by limiting the particle to the  manifold, $M$,this type of potential is defined in terms of the principal curves at each point \cite{17}, meanwhile the kinetic energy is also modified by the metric.
We are particularly interested in circumstances in which there is no physical potential applied.\\

Da Costa proposed a formalism that characterized carriers restricted to a curved surface, by modifying the Schr\"{o}dinger equation and introducing a geometric potential as a function of curvature \cite{12}. This approach is known as the thin-layer quantization method. Thin-layer quantization (a confining potential approach) takes into account all 3D features of carriers. Unlike prior methods, this is a well-defined method that avoids the well-known ambiguity difficulties of operators ordering and adhering to Heisenberg's uncertainty principle. The resulting Schr\"{o}dinger equation for flat surface is the same as  2D Schr\"{o}dinger equation. This explains why this equation is so good at modeling quasi-2D systems.\\

Curvature, on the other hand, has major impacts, such as the geometric potential, which has been the topic of extensive investigations. Both intrinsic and extrinsic curvatures are used to express the geometrical potential \cite{4,6,8,12,14,15,17,35}.\\ 

The limitation is provided as a the result of an external potential that confines the electrons' mobility to a tiny layer of minimal thickness. The Schr\"{o}dinger equation in normal direction (perpendicular to the interface) is found by considering the wave function of electrons as a product of the vertical and tangential wave functions. By separating the dynamic equations of motion in the two directions, the Schr\"{o}dinger equation for the normal direction (perpendicular to the interface) is obtained,
\begin{eqnarray}
\frac{-{{\hbar }^{2}}}{2m}\frac{{{\partial }^{2}}}{\partial {{({{q}^{3}})}^{2}}}{{\chi }_{n}}({{q}^{3}})+V({{q}^{3}}){{\chi }_{n}}={{E}_{n}}{{\chi }_{n}}({{q}^{3}}), 
\label{eq9}
\end{eqnarray}
in which  $V({{q}^{3}})$ stands for the potential that limits the particle to the thin layer, where ${{q}^{3}}$ is the normal coordinate.
Similarly, the Schr\"{o}dinger equation for tangential states is obtained for a spinless particle as follows:
\begin{eqnarray}
	\frac{{{\hbar }^{2}}}{2m}[-\frac{1}{\sqrt{g}}{{\partial }_{\mu }}(\sqrt{g}{{g}^{\mu \nu }}{{\partial }_{\nu }})]{{\chi }_{T}}({{q}_{1}},{{q}_{2}})&+&{{V}_{geometry}}{{\chi }_{T}}({{q}_{1}},{{q}_{2}})\nonumber\\&=&
	E{{\chi }_{T}}({{q}_{1}},{{q}_{2}}),
	\label{eq10}
\end{eqnarray}
where ${{g}^{\mu \nu }}$is the contravariant component of the manifold metric tensor, $g=\det ({{g}_{\mu \nu }})$, and  ${{V}_{geometry}}$ is calculated as a scalar geometric potential that is given by (see Appendix \ref{A} for more information) :
\begin{eqnarray}
{{V}_{geometry}}=-\frac{{{\hbar }^{2}}}{2m}({{H}^{2}}-K),
\label{eq11}
\end{eqnarray}
where $H$ is the mean curvature and K is the Gaussian curvature, which are given using the following relations: $H=\frac{\alpha _{1}^{1}+\alpha _{2}^{2}}{2}$, $K=\alpha _{1}^{1}\alpha _{2}^{2}-\alpha _{1}^{2}\alpha _{2}^{1}$ ; $\alpha _{1}^{1}$ and $\alpha _{2}^{2}$ are principal curvatures of the surface, and $\alpha _{2}^{1}=\alpha _{1}^{2}=0$. Weingarten equations \cite{37} are used to obtain $\alpha $ coefficients (see Appendix \ref{A}). In this approach confining potential is defined as \cite{12,15,32}:
\begin{eqnarray}
V({{q}^{3}})=\begin{cases}
0~~for\left| {{q}^{3}} \right|\le \varepsilon\\
\infty~~otherwise,
\end{cases}
\label{eq12}
\end{eqnarray}
in which the ideal surface is obtained at the limit of $\varepsilon \to 0$. The metric equation defines the Riemannian curved surface that we have considered for ripples as described by Eq. \eqref{eq3}. Using this metric, the modified Schr\"{o}dinger equation can be written as below:
\begin{eqnarray}
\-\frac{{{\hbar }^{2}}}{2m}\{\partial _{r}^{2}+\frac{1}{r}{{\partial }_{r}}+\frac{1}{{{r}^{2}}}\partial _{\theta }^{2}-\alpha f(r)[\partial _{r}^{2}+\frac{1}{r}{{\partial }_{r}}+\nonumber\\
\frac{1}{r}(1-\frac{2{{r}^{2}}}{{{b}^{2}}}){{\partial }_{r}}-\frac{{{r}^{2}}}{{{b}^{4}}}]\}{{\chi }_{T}}(r,\theta )=E{{\chi }_{T}}(r,\theta ).  
\label{eq13}  
\end{eqnarray} 
Equation \eqref{eq13}  has two part. One part is the contribution of the flat space and the other part comes from the space curvature, which we refer to the effective geometric potential. The effective geometric potential has been considered as a perturbation potential to the Schr\"{o}dinger equation
\begin{eqnarray}
  {}^{'}{{V}^{'}}=\frac{{{\hbar }^{2}}}{2m}\alpha f(r)[\partial _{r}^{2}+\frac{1}{r}(1-\frac{2{{r}^{2}}}{{{b}^{2}}}){{\partial }_{r}}-\frac{{{r}^{2}}}{{{b}^{4}}}].                                                   \label{eq14}                  
\end{eqnarray} 
Geometric potential is a non-local potential in real space. It is regarded as a perturbation and Eq. \eqref{eq13} could be solved numerically. In principle, both the correction created by the metric and the correction produced by the geometric shape of the curved surface using the thin-layer method, generate the effective geometric potential represented in Eq. \eqref{eq14}.
\subsection{Green’s function of Schr\"{o}dinger equation in a curved space time}
Equation\eqref{eq13} could be rewritten as
\begin{eqnarray}
	(-\frac{{{\hbar }^{2}}}{2m}{{\nabla }^{2}}+{}^{'}{{V}^{'}}-E){{\chi }_{T}}(r,\theta )=0.   
	\label{eq15}                                                                   
\end{eqnarray}
In curved space, the exact propagator equation is
\begin{eqnarray}
	(-\frac{{{\hbar }^{2}}}{2m}{{\nabla }^{2}}+{}^{'}{{V}^{'}}-E)G(r,\theta ;{r}',{\theta }')=\nonumber\\\frac{1}{r\sqrt{1+\alpha f(r)}}\delta (r-{r}')\delta (\theta -{\theta }').     
	\label{eq16}                                                
\end{eqnarray}
 When the effective potential is removed from Equation \eqref{eq15}, we get the following analytically solvable equation for the perturbation-less propagator
 \begin{eqnarray}
 (-\frac{{{\hbar }^{2}}}{2m}{{\nabla }^{2}}-E)G_{KG}^{0}(r,\theta ;{r}',{\theta }')=\frac{1}{r}\delta (r-{r}')\delta (\theta -{\theta }'). \nonumber\\   
  \label{eq17}                            
 \end{eqnarray}                           The KG subtitle has been used to refer to the Klein-Gordon.
Equation\eqref{eq17} has the following solution  
 \begin{eqnarray}
 	G_{KG}^{0}(\vec{x},\vec{{x}'})=\frac{-i}{4}\frac{2m}{{{\hbar }^{2}}}H_{0}^{(1)}(\sqrt{\frac{2mE}{{{\hbar }^{2}}}}\left| \vec{x}-\vec{{x}'} \right|). 
 	\label{eq18}   
\end{eqnarray}  
To obtain the propagator in the presence of the effective potential, perturbation expansion can be employed. Then the Green's function can be obtained up to the first order of perturbation as follows:
\begin{eqnarray} 
\label{eq19}  
	G(x,{x'})&=&G_0(|{x}-{x'}|)-\frac{1}{2}G_0(|{x}-{x'}|)\alpha f(x)\\&+&\int dx''G_0(|{x}-{x''}|)V(x'')G_0(|{x'}-{x''}|). \nonumber  
\end{eqnarray}
\section{The modified Dirac equation for Fermions on two-dimensional curved surfaces}
It has been assumed that the system consists of a layer of atoms embedded in a flat surface in the case of Dirac mass-less carriers, such as graphene-encapsulated electrons, which defies the uncertainty principle because the particle motion cannot be controlled even in one direction. On this basis, it is expected that graphene is not a perfectly flat surface and it exhibits distortions, which has been experimentally validated \cite{1}.\\ The essential question now is whether the mass-less Dirac equation could be modified, regarding the carriers restriction, despite the effect of ripples in graphene. In the case of Schr\"{o}dinger carriers, we found that the Schr\"{o}dinger equation requires the addition of a purely geometric potential due to curvature, generated by the thin-layer method \cite{10}. We examine that such a term should be included in the mass-less Dirac's equation.\\
Geometric potential was not included in some previous works \cite{6,5}, while it was considered in other works \cite{8,14,15}. In the current paper, we have employed the approach presented in Refs. \cite{6,5}, since it has been shown that the geometrical potential of Dirac equation could be neglected up to the first order approximation \cite{6}. In the non-relativistic scenario, the ripples of a thin layer offer a non-vanishing geometric potential. \\

The advantages of this method over other approaches  mentioned in the previous section could be summarized as follows: This method employs the thin-layer quantization technique, where, in comparison to other methods, this strategy solves the problem of incompatibility with the Heisenberg uncertainty principle. Due to the Heisenberg’s uncertainty principle we can't have zero component of motion in a particular direction.
In other words, there isn't any 2D quantum motion that can occur on a perfectly flat surface. However, in the curved surface the uncertainly principle is valid \cite{15}. This means that the uncertainty principle is not broken when the equation of motion takes into account the third dimension.\\
The curvature can be perceived as a useful gauge field.  Meanwhile, the periodicity of the lattice is disrupted when the distances between atoms change the requirements for utilizing the Bloch function are not satisfied. On this basis, it is obvious that traditional approaches are insufficient, demonstrating the intricacy of graphene ripples.\\ The thin layer approach, which was stated in the preceding section for solving the Schr\"{o}dinger equation, develops a geometric potential by reducing one dimension of the carrier motion from three to two dimensions, confining its mobility to a curved surface. However, applying the same method for Dirac carriers is questionable.\\
To address this question, we utilize the same procedure as described in the Schr\"{o}dinger equation. Appendix \ref{B}  contains the details of this procedure, which was introduced in some of previous works \cite{6,14,15}. The confining process in this method comes from the application of an external real force that eventually reduces the 3D quantum dynamics to two dimensions. In the case of Dirac carriers, we obtain an equation for minor fluctuations in the third dimension, i.e., perpendicular to the plane, which are previously known for non-relativistic carriers and was described in the previous section. Indeed, the downward force of attraction caused by the strain forces created by the curvature keeps the atoms on the surface.\\
The (3 + 1) dimensional Dirac equation can be split into (2 + 1) and (1 + 1) dimensional equations when the perpendicular motion is minimal enough. The quantum dynamics on the surface are described by the (2+1) dimensional equation, while the behavior of the normal component of the wavefunction is described by the (1 + 1) dimensional equation. Heisenberg's principle is not broken in this technique since none of the wave function components are assumed to be zero in the process outlined above \cite{14}.
\\
It is observed that graphene provides an opportunity to study the relationship between quantum mechanics and the theory of gravity, where the empirical confirmations of the relativistic description of quantum motion on a manifold are very important for this subject. This aspect of the issue is very exciting. Several works have been done theoretically in this field.
The equation of Dirac carriers is derived in the same way as the thin layer quantization approach has been performed  \cite{6,14,15}. It was demonstrated that a new term is generated as a geometric potential in the curved surface, similar to the Schr\"{o}dinger case \cite{14,15}. A similar method is used in Ref. \onlinecite{6} to show that geometrical potential could be ignored up to the first order of perturbation of the Dirac equation. Using an approximation of the vertical displacement, details of the approach can be found in a Appendix \ref{B}.
\\
To extend the Dirac equation from flat space-time to curved space-time, the Dirac matrices, ${{\gamma }^{a}}$, must first be substituted by the ${{\gamma }^{\mu }}$ i.e. the curved Dirac matrices. This can be performed by using beins, according to quantum field theory in curved space-time
\begin{eqnarray} 
{{\gamma }^{\mu }}=e_{a}{}^{\mu }{{\gamma }^{a}}. 
\label{eq20}  
\end{eqnarray}
The physical quantities are converted from flat to curved frames and vice versa by using the beins ${e_{a}}^{\mu }$  and their inverse ${e^{a}}_{\mu }$. The equation which may link the metric tensor of the curved space-time to the Minkowski metric has been given by 
\begin{eqnarray} 
{{g}_{\mu \nu }}={{\eta }_{ab}}e^{a}{}_{\mu }e^{b}{}_{\nu }.
\label{eq23}
\end{eqnarray}
 The Dirac equation in flat Minkowski space-time is as follows:
\begin{eqnarray} 
(i\hbar {{\gamma }^{a}}{{\partial }_{a}}-mc)\psi =0  
\label{eq21}.     
\end{eqnarray}
In terms of Minkowski metric, the constant matrices ${{\gamma }^{a}}$ satisfy the following anticommutation relation 
\begin{eqnarray} 
\{{{\gamma }^{a}},{{\gamma }^{b}}\}=2{{\eta }^{ab}}, 
\label{eq22}    
\end{eqnarray}
and similarly for curved space we can write,
\begin{eqnarray} 
\{{{\gamma }^{\mu}},{{\gamma }^{\nu}}\}=2g^{\mu\nu}.
\label{eq58}    
\end{eqnarray}
Because the partial derivative of spinor is not invariant under the coordinate change, i.e. does not transform like a spinor, the partial derivatives should be replaced  with gauge covariant derivative. Spinor's covariant derivative has the following form
\begin{eqnarray} 
{{\nabla }_{\mu }}\psi =({{\partial }_{\mu }}+{{\Omega }_{\mu }})\psi. 
\label{eq24}
\end{eqnarray}
In this case, the spin connection $\Omega_{\mu }$ is
\begin{eqnarray} 
\Omega_{\mu }=\frac{1}{8}{{\omega }_{\mu cd}}[{{\gamma }^{c}},{{\gamma }^{d}}].
\label{eq25}  
\end{eqnarray}
The spin connection ${{\omega }_{\mu }}{{^{c}}_{d}}$ is obtained by 
\begin{eqnarray} 
{{\omega }_{\mu }}{{^{c}}_{d}}={{e}^{c}}_{\nu }{{e}_{d}}^{\sigma }\Gamma _{\sigma \mu }^{\nu }+{{e}^{c}}_{\nu }{{\partial }_{\mu }}{{e}_{d}}^{\nu },
\label{eq26}  
\end{eqnarray}
where $\Gamma _{\sigma \mu }^{\nu }$ is the Christoffel symbol in Levi-Civita connection. The following relation is obtained by substituting Eq. \eqref{eq25} in the Dirac equation of flat space-time, results in mass-less Dirac equation of curved space-time \cite{34}  
\begin{eqnarray} 
i\gamma ^{\mu }\Big(\partial_{\mu }+\frac{1}{8}\omega_{\mu cd}[\gamma^{c},\gamma^{d}])\Psi (t,r,\theta )\Big)=0. 
\label{eq27}
\end{eqnarray}
We recall that, regarding previous works \cite{5,6}, we have realized that the existence of any geometric potential in the modified Dirac equation is not necessary and could be neglected approximately. This means that the different behavior of flat and curved  Dirac-type systems don't merely come from the difference in space-dependent potential. Dynamics of carriers in graphene in low energies is described by the Dirac equation in (2 + 1) dimensional curved space–time.

Equation \eqref{eq23} does not fix $e^{a}{}_{\mu }$ uniquely. Here, we consider the spatial component independently of the time component and reduce the problem to a 2D problem,so the ${{\eta }_{ab}}$ matrix is a 2D unit matrix in this space. There are two obvious solutions :
\begin{eqnarray} 
e^{a}{}_{\mu }&=&\begin{pmatrix}
1 & 0  \\
0 & r  \\
\end{pmatrix},\nonumber\\
e^{a}{}_{\mu }&=&\begin{pmatrix}
\cos \theta  & -r\sin \theta   \\
\sin \theta  & r\cos \theta   \\
\end{pmatrix}.
\label{eq30} 
\end{eqnarray}
The first solution leaves the gamma matrices on the Cartesian plane and induces a constant gauge connection with an obvious zero "associated magnetic field." The second option converts the flat gamma matrices without inducing a gauge connection \cite{5}. We have chosen the second zweibeins as they provide us more degree of freedom in the $\theta$ angle.
\\
The geometric factors of the second choice are
\begin{eqnarray} 
e^{a}{}_{\mu }&=&\begin{pmatrix}
\cos \theta  & -r\sin \theta   \\
\sin \theta  & r\cos \theta   \\
\end{pmatrix},\nonumber\\
e_{a}{}^{\mu }&=&\begin{pmatrix} 
\cos \theta  & \sin \theta   \\
-\frac{\sin \theta }{r} & \frac{\cos \theta }{r} \\
\end{pmatrix}. 
\label{eq35}
\end{eqnarray}

Using the flat metric in the polar coordinate system yields Christoffel coefficients, given in Eqs. \eqref{eqB31} and \eqref{eq35} yields

\begin{eqnarray}
i\hbar {{v}_{f}}({{\gamma }^{0}}\frac{{{\partial }_{t}}}{{{v}_{f}}}+\Gamma (\theta ){{\partial }_{r}}+\frac{{{\partial }_{\theta }}\Gamma (\theta )}{r}{{\partial }_{\theta }})\psi (t,r,\theta )=0 ,
\label{eq37}     
\end{eqnarray}
where $\Gamma (\theta )=\cos \theta {{\gamma }^{1}}+\sin \theta {{\gamma }^{2}}$.
Then the flat Hamiltonian can be written as
\begin{eqnarray}
H{}_{flat}=\hbar {{\nu }_{f}}\begin{pmatrix}
0 & {{e}^{-i\theta}}\big({{\partial }_{r}}-i\frac{{{\partial }_{\theta }}}{r}\big)  \\
{{e}^{i\theta}}\big({{\partial }_{r}}+i\frac{{{\partial }_{\theta }}}{r}\big) & 0  \\
\end{pmatrix}.
\label{eq38}
\end{eqnarray}
The curved space metric in the polar coordinate system, yields Christoffel coefficients and zweibeins (see Appendix \ref{B}).
\\
In this case, the zweibeins and their inverse for the curved surface can be considered as
\begin{eqnarray}
e^{a}{}_{\mu }&=&
\begin{pmatrix}
(1+\alpha f(r))^{\frac{1}{2}}\cos \theta  & -r\sin \theta  \\   
{{(1+\alpha f(r))}^{\frac{1}{2}}}\sin \theta  & r\cos \theta   
\end{pmatrix}, \nonumber  \\
e_{a}{}^{\mu }&=&\begin{pmatrix}
\frac{\cos \theta }{{{(1+\alpha f(r))}^{\frac{1}{2}}}} & \frac{\sin \theta }{{{(1+\alpha f(r))}^{\frac{1}{2}}}}  \\
-\frac{\sin \theta }{r} & \frac{\cos \theta }{r}  
\end{pmatrix}
\label{eq58}
\end{eqnarray}

The following modified Dirac equation is obtained (see Appendix \ref{B}):

\begin{eqnarray}
i\hbar {{v}_{f}}\Big[{{\gamma }^{0}}\frac{{{\partial }_{t}}}{{{v}_{f}}}+\frac{\Gamma (\theta )({{\partial }_{r}}-\frac{1}{2r}+\frac{1}{2r}{{(1+\alpha f(r))}^{\frac{1}{2}}})}{{{(1+\alpha f(r))}^{\frac{1}{2}}}}+\nonumber \\\frac{{{\partial }_{\theta }}\Gamma (\theta )}{r}{{\partial }_{\theta }}\Big]\psi (t,r,\theta )=0,
\label{eq44}
\end{eqnarray} 
in which $\Gamma (\theta )$ is given by
\begin{eqnarray}
\Gamma (\theta )=\cos \theta {{\gamma }^{1}}+\sin \theta {{\gamma }^{2}}.\ \nonumber
\end{eqnarray}
With a little algebra, we can rearrange the above equation into the following form
\begin{eqnarray}
i\hbar {{v}_{f}}\Big[{{\gamma }^{0}}\frac{{{\partial }_{t}}}{{{v}_{f}}}+\Gamma (\theta ){{\partial }_{r}}+\frac{{{\partial }_{\theta }}\Gamma (\theta )}{r}{{\partial }_{\theta }}~~~~~~~~~~~~~~~~~~~~~~~ \\
+\Gamma (\theta )(\frac{1}{{{(1+\alpha f(r))}^{\frac{1}{2}}}}-1)({{\partial }_{r}}-\frac{1}{2r})\Big]\psi (t,r,\theta )=0.\nonumber
\label{eq45}
\end{eqnarray} 
Finally, the Hamiltonian of the curved system is then given as follows:
\begin{widetext}
\begin{eqnarray}
H{}_{curved}=
&&\hbar {{\nu }_{f}}
\begin{pmatrix}
0 & {{e}^{-i\theta}}\Big(\frac{{{\partial }_{r}}}{{{(1+\alpha f(r))}^{\frac{1}{2}}}}-i\frac{{{\partial }_{\theta }}}{r}+ {{A}_{\theta }}\Big)  \\
{{e}^{i\theta}}\Big(\frac{{{\partial }_{r}}}{{{(1+\alpha f(r))}^{\frac{1}{2}}}}+i\frac{{{\partial }_{\theta }}}{r}+ {{A}_{\theta }}\Big) & 0  \\	
\end{pmatrix},
\label{eq46}
\end{eqnarray}
\end{widetext}
in which ${{A}_{\theta }}$ is defined as:
${{A}_{\theta }}=\frac{1}{2r}(1-{{(1+\alpha f)}^{-\frac{1}{2}}})$.
based on the results of Refs. \cite{5,6}.
\subsection{GREEN ’S FUNCTIONS OF CURVED DIRAC EQUATION}
In curved space, the general mass-less Dirac equation is: $
i\hbar {{\gamma }^{\mu }}({{\partial }_{\mu }}+{{\Omega }_{\mu }})\psi =0$, hence Green's functions meet the following equation
\begin{eqnarray}
i\hbar {{\gamma }^{\mu }}({{\partial }_{\mu }}+{{\Omega }_{\mu }})G(x,{x}')=\frac{1}{\sqrt{-g}}\delta (x-{x}').
\label{eq47}
\end{eqnarray}
As shown in previous work \cite{5}, it can be seen that:\\ $\alpha \ll 1$. Accordingly:
\\
$\frac{1}{r{{(1+\alpha f(r))}^{\frac{1}{2}}}}\approx \frac{1}{r}(1-\frac{1}{2}\alpha f(r)).$
By substituting this relation in the above equation, one can obtain
\begin{eqnarray}
i\hbar {{\gamma }^{\mu }}({{\partial }_{\mu }}+{{\Omega }_{\mu }}){{G}_{D}}(x,{x}')&+&\frac{\alpha f(r)}{2r}\delta (x-{x}')\nonumber \\&=&\frac{1}{r}\delta (x-{x}').
\label{eq48}	
\end{eqnarray}
We get the following relation by replacing the preceding equations for Dirac and spin connection matrices: 
\begin{widetext}
	\begin{eqnarray}
	i\hbar {{v}_{f}}\Big[{{\gamma }^{0}}\frac{{{\partial }_{t}}}{{{v}_{f}}}+\Gamma (\theta ){{\partial }_{r}}+\frac{{{\partial }_{\theta }}\Gamma (\theta )}{r}{{\partial }_{\theta }}+\Gamma (\theta )(\frac{1}{{{(1+\alpha f(r))}^{\frac{1}{2}}}}-1)({{\partial }_{r}}-\frac{1}{2r})\Big]{{G}_{D}}(\vec{x},\vec{{x}'})+\frac{1}{2r}\alpha f(r)\delta (x-{x}')=\frac{1}{r}\delta (x-{x}'),\nonumber \\
	\label{eq49}
	\end{eqnarray}
\end{widetext}
where $G_D$ is the curved surface Green's function. The last term inside the brackets of the above relation acts as an effective potential caused by the surface curvature. Up to the first order in $\alpha $, we get
\begin{eqnarray}
{}^{'}{{V}^{'}}=i\hbar {{v}_{f}}\Gamma (\theta )(\frac{1}{{{(1+\alpha f(r))}^{\frac{1}{2}}}}-1)({{\partial }_{r}}-\frac{1}{2r}).\nonumber
\end{eqnarray}
Using the approximation $\alpha\ll 1$, the following expression is obtained:
${}^{'}{V}^{'}\approx -i\hbar {{v}_{f}}\Gamma (\theta )(\frac{1}{2}\alpha f(r))({\partial_{r}}-\frac{1}{2r})$.
If $G_{D}^{0}(\vec{x},\vec{{x}'})$ is a Green's function, that satisfies the following differential equation:
\begin{eqnarray}
[{{\gamma }^{0}}E+i\hbar {{v}_{f}}(\Gamma (\theta ){{\partial }_{r}}+\frac{{{\partial }_{\theta }}\Gamma (\theta )}{r}{{\partial }_{\theta }})]G_{D}^{0}(\vec{x},\vec{{x}'})\nonumber \\
=\frac{1}{r}\delta (r-{r}')\delta (\theta -{\theta }').  
\label{eq50}           
\end{eqnarray}
Using the relations, $[\Gamma (\theta )]^{2}=[{{\partial }_{\theta }}\Gamma (\theta )]^{2}=-[{\gamma }^{0}]^{2}=-1$, $\{{\gamma}^{0},\Gamma (\theta )\}=\{{\gamma }^{0},{{\partial }_{\theta }}\Gamma (\theta )\}=\{\Gamma (\theta ),{{\partial }_{\theta }}\Gamma (\theta )\}=0$ , 
it can be demonstrated that multiplying this differential operator by itself from the left hand side leads to the Klein-Gordon equation:
\begin{eqnarray}
[{{E}^{2}}+{{(\hbar {{v}_{f}})}^{2}}(\partial _{r}^{2}+\frac{1}{r}{{\partial }_{r}}+\frac{1}{{{r}^{2}}}\partial _{_{\theta }}^{2})]G_{KG}^{0}(\vec{x},\vec{{x}'})\nonumber \\=\frac{1}{r}\delta (r-{r}')\delta (\theta -{\theta }')
\label{eq51}.
\end{eqnarray}  
Comparing, Eqs. \eqref{eq50} and \eqref{eq51}, one can obtain,
\begin{eqnarray}
[{{\gamma }^{0}}E+i\hbar {{v}_{f}}(\Gamma (\theta ){{\partial }_{r}}+\frac{{{\partial }_{\theta }}\Gamma (\theta )}{r}{{\partial }_{\theta }})]G_{KG}^{0}(\vec{x},\vec{{x}'})\nonumber \\=
G_{D}^{0}(\vec{x},\vec{{x}'}).
\label{eq53}
\end{eqnarray}
As a result, it is obtained from Eq. \eqref{eq53}
\begin{widetext}
	\begin{eqnarray}
	G_{D}^{0}(| \vec{x}-\vec{{x}'} |,\frac{E}{\hbar {{v}_{f}}})=\frac{-i}{4}\frac{E}{{{(\hbar {{v}_{f}})}^{2}}}{{\gamma }^{0}}{{H}_{0}}(\frac{E}{\hbar {{v}_{f}}}| \vec{x}-\vec{{x}'} |)-\frac{E}{{{(\hbar {{v}_{f}})}^{2}}}\frac{1}{4| \vec{x}-\vec{{x}'} |}{{H}_{1}}(\frac{E}{\hbar {{v}_{f}}}| \vec{x}-\vec{{x}'} |)[r\Gamma (\theta )-{r}'\Gamma ({\theta }')].
	\label{eq54}
	\end{eqnarray}
\end{widetext}

The first-order correction of ${{\text{G}}_{D}}$, can be obtained by a perturbation expansion in which the Green's propagator in the presence of effective potential is given with \eqref{eq19}.\\
\section{Results and  Discussion}
\begin{figure}[htbp]
	
	\centering
	
	\includegraphics[width=8cm,height=10cm]{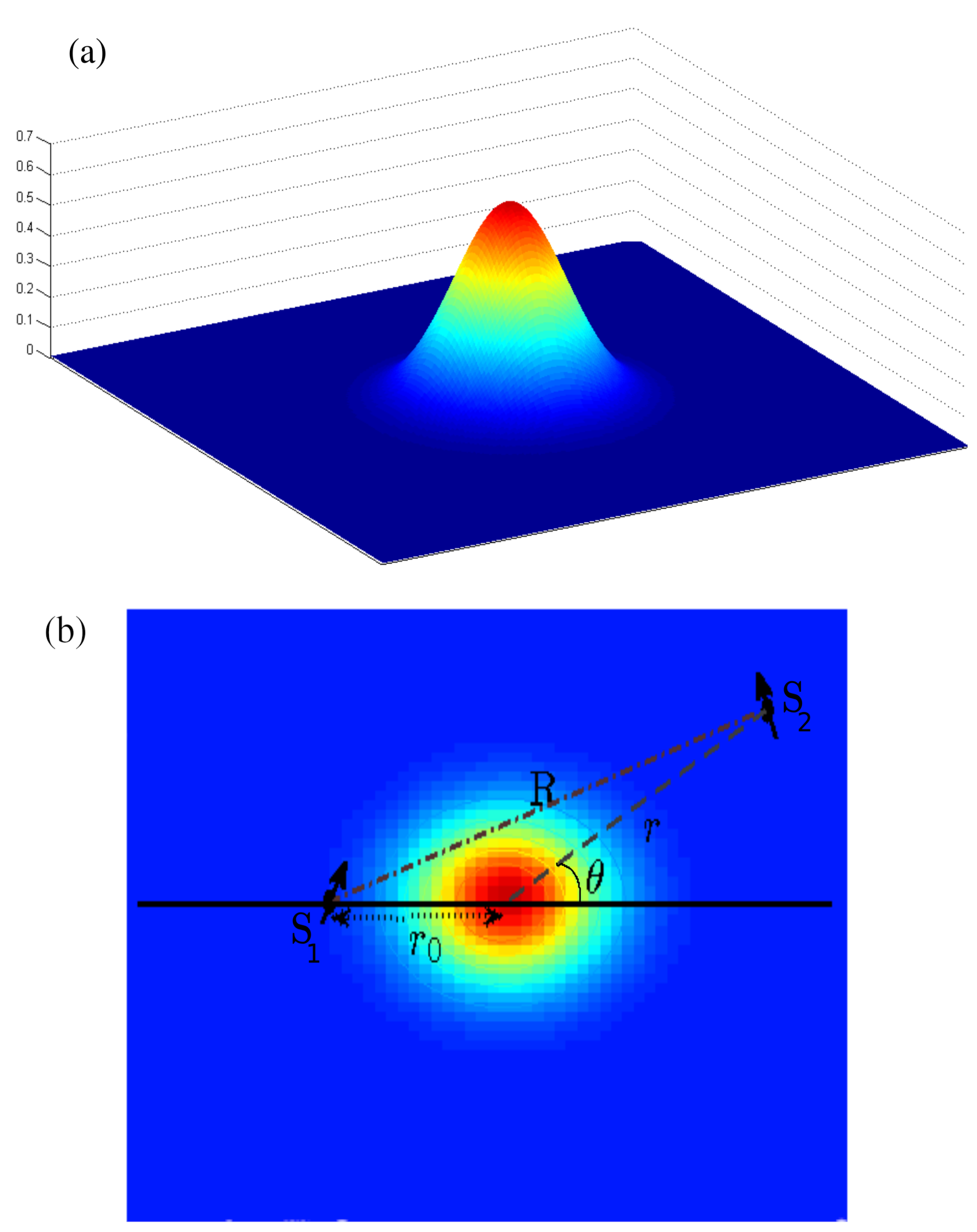}
	\caption{\texttt (Color online) (a) Side view of the a Gaussian bump. (b) Top view of the sample where, the position of magnetic impurities, $S_1$ and $S_2$, inside the system is given by $(r_0, \pi)$ and $(r,\theta)$ respectively. $R$ is the direct distance between the impurities. The bump center has been considered as the polar coordinate origin.}
	\label{fig:2}
	
\end{figure}

The RKKY interaction strength and the LDOS are calculated using Eqs. \eqref{eq7} and \eqref{eq8}. The results are shown for both  flat and curved surfaces. The RKKY interaction and LDOS is then estimated for two magnetic impurities in curved surfaces, regardless of the material's intrinsic structure, as illustrated in Fig. \ref{fig:2}.  The curved surface in question has been regarded as the same profile for both Schr\"{o}dinger and Dirac carriers where the Fermi energy is assumed to be $E_F=0.1$(eV) for both samples.  
\\
The influence of the curved surface on the RKKY interaction could then be captured by Green's functions. We have estimated the spin susceptibility and LDOS for two flat and curved surfaces containing magnetic impurities using Eqs. \eqref{eq5} and \eqref{eq8}.
 \\ 
 
The bump is located in the center of the surface and one of the impurities located on the x axis, the distance between this fixed impurity and the bump center is indicated by ${{r}_{0}}$. The second impurity is considered at $(r,\theta )$ as measured from the bump center in the polar coordinate as shown in Fig. \ref{fig:2}.\\

\begin{figure*}[htbp]
	
	\centering
	
	\includegraphics[width=10cm,height=8cm]{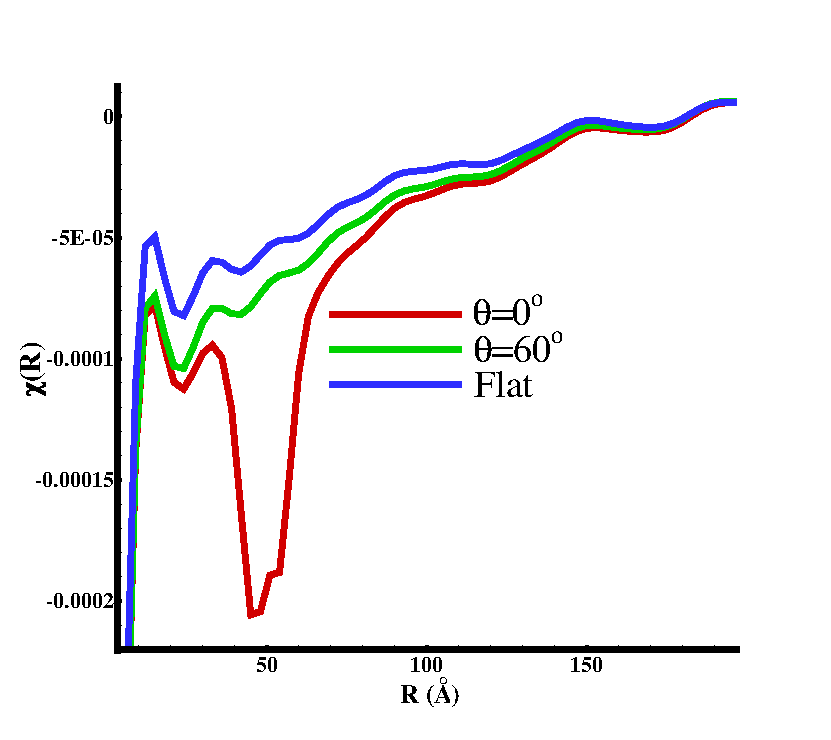}
	\caption{\texttt (Color online) The susceptibility in two different curved states in comparison with that of  the flat case susceptibility for Schr\"{o}dinger carriers. The susceptibility is given in the unit of $eV^{-1}\AA^{-4}$. The susceptibility has been obtained for $b = 50\AA$ and ${{r}_{0}} = 50\AA$ at different $\theta $ angles of ${{0}^{{}^\circ }}$ and ${{60}^{{}^\circ }}$ degrees where the Fermi energy is $E_F=0.1$(eV). It is clear that the absolute value of the RKKY interaction is increased at the bumped area.  }
	\label{fig:3}
	
\end{figure*}

Spin susceptibility has been computed for flat and curved surfaces in several directions (at different polar angles, $\theta $) as shown in Figs. \ref{fig:3} and \ref{fig:5} .\\ 

\begin{figure*}[htbp]
	
	\centering
	
	\includegraphics[width=10cm,height=8cm]{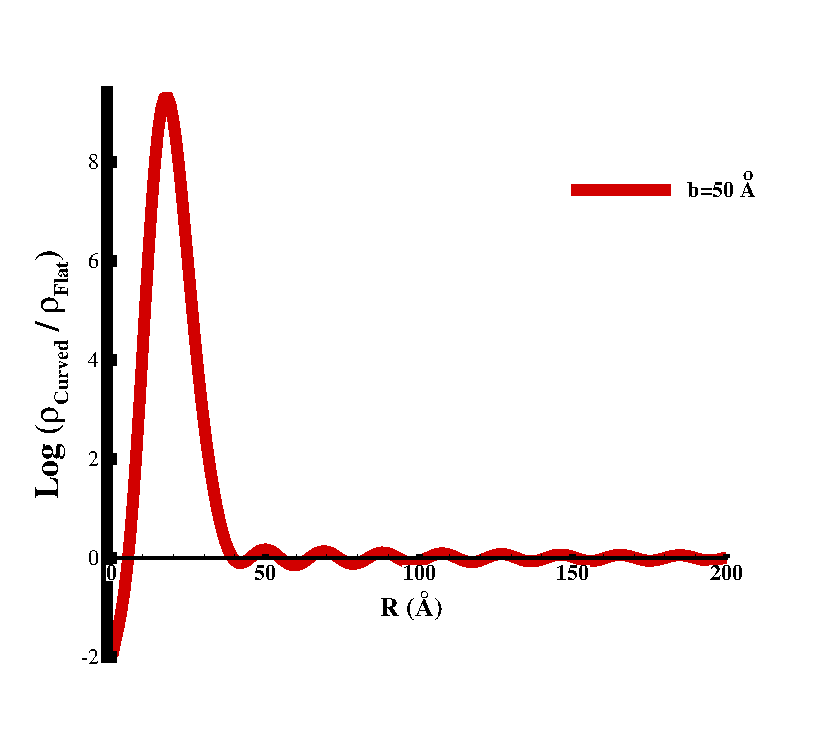}
	\caption{\texttt The ratio of the curved surface  local density to that of the flat surface for Schr\"{o}dinger carriers at $E_F=0.1$(eV). It can be inferred that the LDOS is increased significantly at the bump area.}
	\label{fig:4}
\end{figure*}

\begin{figure*}[htbp]
	
	\centering
	
	\includegraphics[width=10cm,height=8cm]{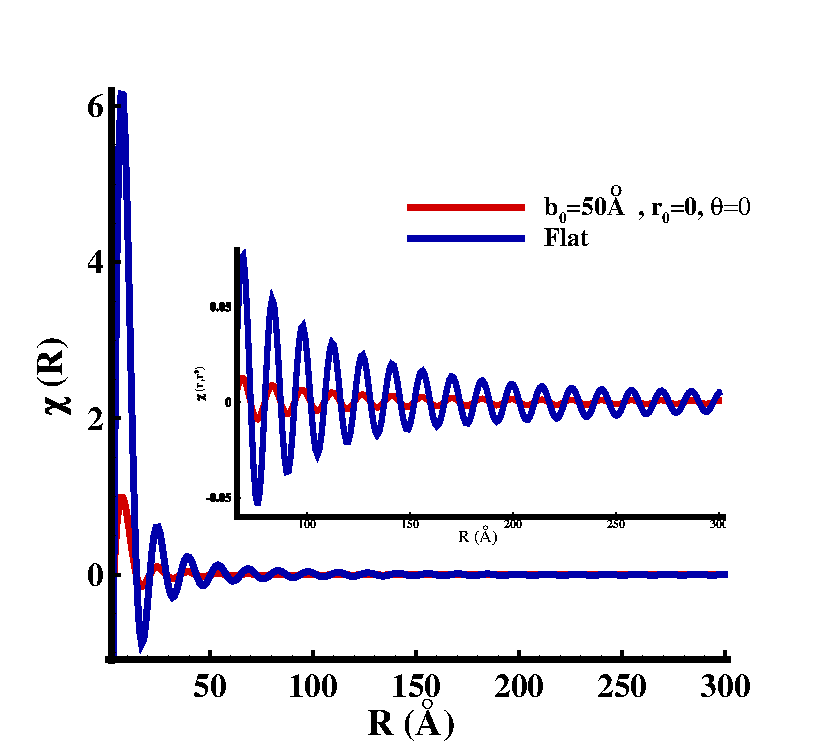}
	\caption{\texttt (Color online) Susceptibility of the flat sample compared with the susceptibility of the curved system for Dirac carriers where the Fermi energy is $E_F=0.1$(eV). Both of them are given in the unit of $eV^{-1}\AA^{-4}$. However, to obtain a more clear illustrative plot, flat and curved sample susceptibilities have been depicted in two different scales.  The scale of the curved system is $10^{-8}$, while the scale of the flat sample is given to be $10^{-5}$. This figure shows that the strength of the RKKY interaction is decreased substantially in the presence of the Dirac carriers. }
	\label{fig:5}
\end{figure*}

\begin{figure*}[htbp]
	
	\centering
	
	\includegraphics[width=10cm,height=8cm]{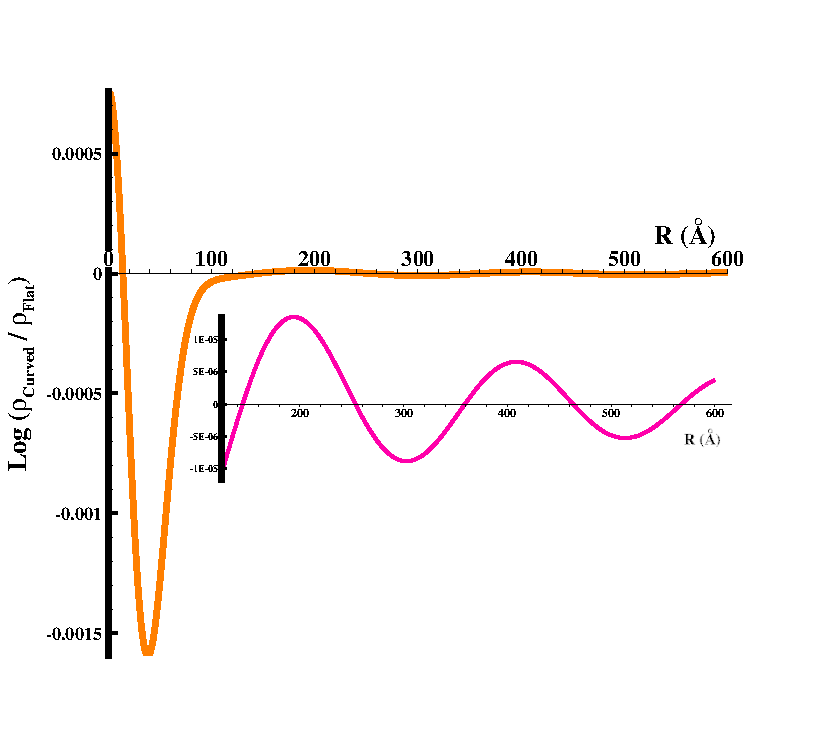}
	\caption{\texttt (Color online) The ratio of the curved surface local density to that of the flat surface for Dirac carriers at $E_F=0.1$(eV). It is clear that the LDOS is decreased significantly at the bump area. The inset shows focused oscillations of density ratios beyond the bump region  after $100 \AA$. }
	\label{fig:6}
\end{figure*}

Fig. \ref{fig:4} compares the local density of curved and flat surfaces for Schr\"{o}dinger carriers, while Fig. \ref{fig:6} compares the local density of curved and plane surfaces for Dirac carriers. It has been shown that the behavior of the local density of states (LDOS) is completely different for Dirac and Schr\"{o}dinger carriers. Comparing Figs. \ref{fig:3} and \ref{fig:5}, it can be observed that the strength of RKKY interaction in two types of carriers also behave reversely. While the absolute value of the RKKY interaction increases in the presence of the Gaussian bump for Schr\"{o}dinger carriers, it decreases for Dirac particles.
\\
It can be inferred that there is an interesting correlation between the intensity of the RKKY interaction and the local density of the states. The LDOS of Schr\"{o}dinger carriers reveals a relative rise around the bump of the curved area. This can be compatible with the increased strength of the RKKY interaction; meanwhile the LDOS of Dirac carriers displays a reduction around the top of the bump which can be understood by the reduction of the RKKY strength. This correlation between the electron density and RKKY interaction strength can easily be inferred regarding the fact that the RKKY interaction between localized impurities is mediated by the conducting electrons so areas with low electron densities could not provide strong RKKY interaction.
\\
Figures \ref{fig:3}, \ref{fig:5} show the strength of the RKKY interaction for two $\theta =0{}^\circ $ and $\theta =60{}^\circ $ directions. The change of the direction has no effect on the RKKY interaction in the case of Dirac carriers, however, in the case of Schr\"{o}dinger carriers, this change results in a meaningful difference as shown in Fig 3.\\
The strength of the RKKY interaction shows a very small change by varying the distance of the impurity from the bump center, ${{r}_{0}}$.  Both the RKKY interaction and LDOS increase at the curved region for Schr\"{o}dinger carriers however, these parameters decrease at the curved position for Dirac carriers.

The effect of curvature is simply perceived in the Schr\"{o}dinger case as an effective potential that can disrupt the distribution of conducting electrons and LDOS  indicates that this potential results in  the accumulation of electrons in the bump region. As the number of conducting electrons transmitting the spin interaction is increased, it is expected that the interaction strength is increased dramatically compared to the flat case.\\
On the other hand, the curvature is seen by Dirac carriers as an effective potential, causing a redistribution of conduction electrons. In this case, the coupling of spin connection and momentum in the Dirac equation allows us to consider it as a vector potential, which  may explain the different behavior of Dirac and Schr\"{o}dinger carriers. One part of the effective potential of Dirac carriers that is directly tied to the spin connection results in generation of a magnetic field Eq. \eqref{eq57}. In this way the bump effect can be replaced with a magnetic field. The vector potential, according to Hamilton Eq.\eqref{eq46}, can be written as follows \onlinecite{5,15}:
\begin{eqnarray}
\label{eq55}
{{A}_{\theta }}&=&\frac{{{\Omega }_{\theta }}}{2r}\\&=&
\frac{1-{{(1+\alpha f(r))}^{-\frac{1}{2}}}}{2r} \nonumber \\&\approx&\ \frac{\alpha f(r)}{4r}.
\nonumber  
\end{eqnarray} 
Then the magnetic field is given by:
\begin{eqnarray}
\label{eq56}	
{{B}_{z}}&=&-\frac{1}{r}{{\partial }_{r}}(r{{A}_{\theta }})  \\
&=&\frac{1}{4r}\frac{\alpha {f}''(r)}{{{(1+\alpha f(r))}^{\frac{3}{2}}}}.\nonumber
\end{eqnarray}
The magnetic vector potential can be used to express the effective potential:
\begin{eqnarray}
\label{eq57}
{}^{'}{{V}^{'}}\approx -i\hbar {{v}_{f}}\Gamma (\theta ){{A}_{\theta }}(2r{{\partial }_{r}}-1).
\end{eqnarray}
The magnetic field strength in the bump area is estimated to be between 0.5 and 2-3 Tesla \cite{5,18}. This magnetic field formed by the curvature is responsible for a large reduction in the intensity of the RKKY interaction, even at  positions away from the bump. Furthermore, results of the LDOS are also in agreement with earlier findings \cite{5}.

The intensity of the RKKY interaction does not change significantly with respect to the position vector of the impurities measured from the center of bump. When two-dimensional electrons are exposed to a magnetic field, one of the most well-known quantum effects in solids is displayed. Electronic wave functions are mainly limited to cyclotron circles with discrete energy levels called Landau surfaces. Graphene electrons, which are mass-less quasi-particle fermions, collect on orbits whose radius and center are modified by the inverse of the magnetic field. Because of the presence of this high magnetic field in the bump area, accumulation occurs in circles in which the centers are located near the top of the bump with extremely small radii. Therefore, this magnetic field results in dramatically increasing charge accumulation at the top vertex and reducing of the local density of electronic states in the bump area. Accordingly, in the case of a curved surface, this determines the intensity of the RKKY interaction at different positions so the positions with low electron density show weak RKKY interaction and vice versa.\\

For a given magnetic field $B$, the distance between the center of Landau orbits and the center of the bump is given by $\Delta \approx \frac{\hbar k}{eB}$. The radius of landau orbits is obtained by $r\approx 2\sqrt{\frac{\hbar }{eB}}$ \cite{36}( Appendix C contains the details of the calculation). In the case of the Dirac equation, we can expect a reduction in RKKY interaction as a result of the decrement of electron density outside the bump tip. In the case of the Schr\"{o}dinger carriers, the effective potential regulates the density of states in the bump area, increasing the density of states in the bump area and therefore amplifying the RKKY interaction. As a result, the different behavior of Dirac and Schr\"{o}dinger carriers revealed. 
\\
Finally, it should also be noted that the numerical results may depend on the value of the Fermi energy where the real-space period of oscillations changes by Fermi energy. However, qualitative results i.e. the different behavior of the Dirac and Schr\"{o}dinger carriers don't depend on Fermi energy. 
\section{Conclusion}
On 2D curved surfaces, we estimated the RKKY interaction in two different cases. In the first case, carriers obey the Schr\"{o}dinger equation and the second case corresponds to the carriers in which their dynamics are determined by the Dirac equation.  We used the thin-layer quantization method, also known as the Costa method, to modify the equations in both cases since this approach preserves the Heisenberg uncertainty principle. In the case of Schr\"{o}dinger carriers, a geometric potential is added to the Hamiltonian. However, in the case of Dirac carriers, because of the spinor nature of  the wave function, the covariant derivative is changed in terms of zweibeins and it can be shown that the geometric potential can be neglected up to the first order approximation \cite{6}.
\\
The curvature of the surface could be considered as a source of an effective magnetic field for Dirac carriers, and the presence of this magnetic field results in completely different behavior of Schr\"{o}dinger and Dirac carriers. 
\\
There is an effective geometric potential for both of the carriers studied. The Costa method for Schr\"{o}dinger carriers leads to a potential that is proportional to the square of the distance between the two impurities. However, in the case of Dirac carriers this potential can be neglected, up to first order perturbation. However, Dirac carriers experience an effective magnetic field associated with spin connections that mainly determines the difference between the behavior of these two types of carriers. In general, an extra effective geometric potential corrects LDOS and the RKKY interaction in comparison with the flat surface. Schr\"{o}dinger carriers have a higher LDOS around the bump area, which is followed by a higher RKKY interaction, whereas Dirac carriers have a lower LDOS and a lower RKKY interaction. In the case of Dirac carriers, the generated magnetic field can disrupt the homogeneous spatial distribution of states, and the focus radius of the electron states can rise or decrease, depending on the intensity of the magnetic field. This radius is inversely proportional to the bump's curvature. As the effective magnetic field grows close to the bump, LDOS and the RKKY interaction fall in the bump area, where the magnetic field is stronger.

\appendix
\section{The (2 + 1) Curved Schr\"{o}dinger Equation in Polar Coordinates}
\label{A}
The Costa technique  is used to study the dynamics of non-relativistic carriers constrained to the surface, which is explored briefly here \cite{12}. In curvilinear coordinates $({{q}^{1}},{{q}^{2}},{{q}^{3}})$, the position of an electron limited to the vicinity of the surface is expressed as:
\begin{eqnarray}
\vec{R}({{q}^{1}},{{q}^{2}},{{q}^{3}})=\vec{r}({{q}^{1}},{{q}^{2}})+{{q}^{3}}{{\hat{e}}_{3}}({{q}^{1}},{{q}^{2}}),
\label{eqA1}
\end{eqnarray}

\begin{figure}[htbp]
	
	\centering
	
	\includegraphics[width=8cm,height=5cm]{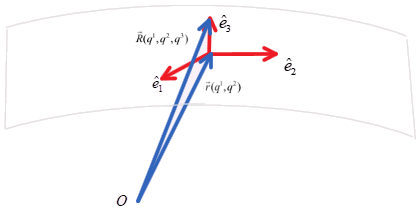}
	\caption{\texttt (Color online) View of the curved surface and vectors representing the points in its vicinity   }
	\label{fig:7}
\end{figure} 
where $\vec{r}({{q}^{1}},{{q}^{2}})$ is the curved surface's parametric equation and $\hat{e}{}_{3}({{q}^{1}},{{q}^{2}})$ is the surface's unit normal vector at $({{q}^{1}},{{q}^{2}})$. Because the derivative of the surface normal vector is tangent to this surface, we can write:
\begin{eqnarray}
{{\partial }_{i}}{{\hat{e}}_{3}}=\alpha _{i}^{j}{{\partial }_{j}}\vec{r}({{q}^{1}},{{q}^{2}}), ~~~~  {{\partial }_{i}}\equiv \frac{\partial }{\partial {{q}^{i}}}
\label{eqA2}
\end{eqnarray}
Weingarten equations are used to obtain $\alpha _{i}^{j}$. The following results are obtained: 
\begin{eqnarray}
{{\partial }_{m}}\vec{R}=(\delta _{m}^{n}+{{q}^{3}}\alpha _{m}^{n}){{\partial }_{n}}\vec{r}, ~~ {{\partial }_{3}}\vec{R}={{\hat{e}}_{3}},
\label{eqA3}                
\end{eqnarray}
where  $m, ~n=1,~2$.    
Next, the metrics between the two coordinate frames shown in Fig. \ref{fig:7} are now obtained: \\
\begin{eqnarray}
G&=&\det ({{G}_{ij}}({{q}^{{1}}},{{q}^{2}},{{q}^{3}}))\nonumber \\  
&=&\det ({{\partial }_{i}}\vec{R}({{q}^{{1}}},{{q}^{2}},{{q}^{3}}).{{\partial }_{j}}\vec{R}({{q}^{{1}}},{{q}^{2}},{{q}^{3}}))\nonumber \\ 
&=&{{\Lambda }^{2}}\det ({{\partial }_{i}}\vec{r}({{q}^{1}},{{q}^{2}}).{{\partial }_{j}}\vec{r}({{q}^{1}},{{q}^{2}}))\nonumber \\ 
& =&{{\Lambda }^{2}}\det ({{g}_{ij}}({{q}^{1}},{{q}^{2}},0))={{\Lambda }^{2}}g. 
\label{eqA4}
\end{eqnarray}
Then we have: $\sqrt{G}=\Lambda \sqrt{g}$.
As illustrated in Fig. \ref{fig:7}, ${{G}_{ij}}({{q}^{1}},{{q}^{2}},{{q}^{3}})$ is the metric of the points on the curved surface with respect to the coordinate frame, and ${{g}_{ij}}({{q}^{1}},{{q}^{2}},0)$ is the metric of the curved surface relative to the local frame on the surface, Fig. \ref{fig:7}. We can obtain the following relationship using some algebra:  
\begin{eqnarray}
\Lambda =1+{{q}^{3}}Tr(\alpha _{m}^{n})+{{({{q}^{3}})}^{2}}Det(\alpha _{m}^{n}).
\label{eqA5}
\end{eqnarray} 
Within  the Costa approach, a potential is defined by following equation \cite{12}:
\begin{eqnarray}
V({{q}^{3}})=\begin{cases}
0~~~for\left| {{q}^{3}} \right|\le \varepsilon\\

\infty ~~~otherwise
\end{cases}
\label{eqA6}
\end{eqnarray}
where $\varepsilon$ is the thickness of the curved surface $S$. 
\\
Now the Schr\"{o}dinger equation is to be modified for the curved surface. To this end, first it is better to obtain the relation between two wave functions in two coordinate systems and then the condition of  preserving total number of particles is applied. If $\Psi ({{q}^{1}},{{q}^{2}},{{q}^{3}})$ is assumed to be the wave function in the global coordinate  system and $\phi ({{q}^{1}},{{q}^{2}},{{q}^{3}})$ is that of in local coordinate system. then:
\begin{eqnarray}
\int{{{d}^{3}}x\sqrt{G}{{\Psi }^{\dagger }}\Psi =\int{{{d}^{3}}x\sqrt{g}{{\Phi }^{\dagger }}}}\Phi.
\label{eqA7}
\end{eqnarray}
Using the Eq. \eqref{eqA4}, following relation is achieved 
\begin{eqnarray}
\Phi (t,{{q}^{{1}}},{{q}^{2}},{{q}^{3}})=\sqrt{\Lambda }\Psi (t,{{q}^{{1}}},{{q}^{2}},{{q}^{3}}).
\label{eqA8}
\end{eqnarray}
In the following, by replacing the electron wave function in terms of local coordinates at the Schr\"{o}dinger equation,
\begin{eqnarray}
\hat{H}\Psi (t,{{q}^{{1}}},{{q}^{2}},{{q}^{3}})=i\hbar \frac{\partial }{\partial t}\Psi (t,{{q}^{{1}}},{{q}^{2}},{{q}^{3}})
\label{eqA9}
\end{eqnarray} 
and by inserting Eq. \eqref{eqA8} in Eq. \eqref{eqA9} one can obtain:
\begin{eqnarray}
i\hbar \frac{\partial }{\partial t}\frac{\Phi (t,{{q}^{1}},{{q}^{2}},{{q}^{3}})}{\sqrt{\Lambda }}=[-\frac{{{\hbar }^{2}}}{2m}\frac{1}{\sqrt{G}}\frac{\partial }{\partial {{q}^{i}}}(\sqrt{G}{{G}^{ij}}\frac{\partial }{\partial {{q}^{j}}})\nonumber \\ +V({{q}^{3}})]\frac{\Phi (t,{{q}^{{1}}},{{q}^{2}},{{q}^{3}})}{\sqrt{\Lambda }},i,j=1,2,3~~~~~~~~~~~~~~~~~~~~~~~~
\label{eqA10}
\end{eqnarray}

Using Eq. \eqref{eqA5}, we get
\begin{widetext}
	\begin{eqnarray}
	i\hbar \frac{\partial }{\partial t}\Phi (t,{{q}^{{1}}},{{q}^{2}},{{q}^{3}})&=&-\frac{{{\hbar }^{2}}}{2m}[\frac{1}{\sqrt{g}}\frac{\partial }{\partial {{q}^{i}}}(\sqrt{g}{{G}^{ij}}\frac{\partial }{\partial {{q}^{j}}})+\frac{{{\partial }^{2}}}{\partial {{({{q}^{3}})}^{2}}}+\frac{1}{4{{\Lambda }^{2}}}{{(\frac{\partial \Lambda }{\partial {{q}^{3}}})}^{2}}-\frac{1}{2\Lambda }(\frac{{{\partial }^{2}}\Lambda }{\partial {{({{q}^{3}})}^{2}}})]+V({{q}^{3}})\Phi (t,{{q}^{{1}}},{{q}^{2}},{{q}^{3}}).
	\nonumber\\ i,j&=&1,2
	\label{eqA11}
	\end{eqnarray}
	The wave equation is divided into two parts. The first part depends on the surface coordinates, and the second part depends on the normal coordinate 
	\begin{eqnarray}
	\Phi (t,{{q}^{{1}}},{{q}^{2}},{{q}^{3}})={{\chi }_{T}}(t,{{q}^{1}},{{q}^{2}}){{\chi }_{N}}(t,{{q}^{3}}).
	\label{eqA12}
	\end{eqnarray}
	The portion of the wave function that reflects behavior along the perpendicular direction is just a function of ${q}^{3}$ which is constrained. We get the following equations within the surface and perpendicular to the surface, respectively
	\begin{eqnarray}
	i\hbar \frac{\partial }{\partial t}{{\chi }_{T}}(t,{{q}^{{1}}},{{q}^{2}})=-\frac{{{\hbar }^{2}}}{2m}[\frac{1}{\sqrt{g}}\frac{\partial }{\partial {{q}^{i}}}(\sqrt{g}{{G}^{ij}}\frac{\partial }{\partial {{q}^{j}}})+\frac{1}{4{{\Lambda }^{2}}}{{(\frac{\partial \Lambda }{\partial {{q}^{3}}})}^{2}}-\frac{1}{2\Lambda }(\frac{{{\partial }^{2}}\Lambda }{\partial {{({{q}^{3}})}^{2}}})]{{\chi }_{T}}(t,{{q}^{{1}}},{{q}^{2}}).
	\label{eqA13}
	\end{eqnarray}
	Additionally, one can obtain following relation for the vertical component
	\begin{eqnarray}
	i\hbar \frac{\partial }{\partial t}{{\chi }_{N}}({{q}^{3}},t)=-\frac{{{\hbar }^{2}}}{2m}\frac{{{\partial }^{2}}}{\partial {{({{q}^{3}})}^{2}}}{{\chi }_{N}}({{q}^{3}},t)+V({{q}^{3}}){{\chi }_{N}}({{q}^{3}},t).
	\label{eqA14}
	\end{eqnarray}
	If ${{q}^{3}}\approx \varepsilon $ we will have $V({{q}^{3}})=\infty $ and ${{\chi }_{N}}({{q}^{3}},t)=0$ , on the other hand ${{q}^{3}}=0$, indicating that the particle is on the surface, and we obtain: $\left| {{\chi }_{N}}({{q}^{3}},t) \right|=1$.
	From Eq. \eqref{eqA5} it is found that $\frac{\partial \Lambda }{\partial {{q}^{3}}}=Tr\alpha _{m}^{n}+2{{q}^{3}}Det(\alpha _{m}^{n})$ and $\frac{{{\partial }^{2}}\Lambda }{\partial {{({{q}^{3}})}^{2}}}=2Det(\alpha _{m}^{n} )$.
	Using confinement on the surface: ${{q}^{3}}\to 0$ then $\Lambda \to 1$  and therefore $\Phi (t,{{q}^{{1}}},{{q}^{2}},{{q}^{3}})\to {{\chi }_{T}}(t,{{q}^{{1}}},{{q}^{2}})$ which results in  $\frac{\partial \Phi }{\partial {{q}^{3}}}\to 0$.
	Accordingly, the Schr\"{o}dinger equation converts to
	\begin{eqnarray}
	E{{\chi }_{T}}(t,{{q}^{{1}}},{{q}^{2}})=-\frac{{{\hbar }^{2}}}{2m}\{\frac{1}{\sqrt{g}}\frac{\partial }{\partial {{q}^{i}}}(\sqrt{g}{{g}^{ij}}\frac{\partial }{\partial {{q}^{j}}})+\frac{1}{4}[{{(Tr\alpha _{m}^{n})}^{2}}-4Det(\alpha _{m}^{n})]\}{{\chi }_{T}}(t,{{q}^{{1}}},{{q}^{2}}).
	\label{eqA15} 
	\end{eqnarray}
	By inserting the metric of the curved surface described in Eq .\eqref{eq3}, we get
	
	\begin{eqnarray}
	-\frac{{{\hbar }^{2}}}{2m}\{[\frac{1}{1+\alpha f(r)}\partial _{r}^{2}+\frac{1}{r(1+\alpha f(r))}{{\partial }_{r}}+\alpha f(r)\frac{\frac{2{{r}^{2}}}{{{b}^{2}}}-1}{r{{(1+\alpha f(r))}^{2}}}{{\partial }_{r}}+\frac{1}{{{r}^{2}}}\partial _{\theta }^{2}]+\alpha f(r)\frac{\frac{{{r}^{2}}}{{{b}^{4}}}{{(1+\frac{2{{z}^{2}}}{{{b}^{2}}})}^{2}}}{{{(1+\alpha f(r))}^{3}}}\}{{\chi }_{T}}(r,\theta )\nonumber \\=E{{\chi }_{T}}(r,\theta ).
	\label{eqA16}
	\end{eqnarray}	
	Because it is assumed that: $\alpha \ll 1$, one can write: ${{(1+\alpha f(r))}^{-n}}\approx 1-n\alpha f(r)$ up to the first order of $\alpha f(r)$.
	On the other hand, we know that: $z\le A$; as a result, $\frac{{{z}^{2}}}{{{b}^{2}}}\le \frac{{{A}^{2}}}{{{b}^{2}}}=\alpha $, and so, we can get \begin{eqnarray}
	-\frac{{{\hbar }^{2}}}{2m}\{\partial _{r}^{2}+\frac{1}{r}{{\partial }_{r}}+\frac{1}{{{r}^{2}}}\partial _{\theta }^{2}-\alpha f(r)[\partial _{r}^{2}+\frac{1}{r}{{\partial }_{r}}+\frac{1}{r}(1-\frac{2{{r}^{2}}}{{{b}^{2}}}){{\partial }_{r}}-\frac{{{r}^{2}}}{{{b}^{4}}}]\}{{\chi }_{T}}(r,\theta )=E{{\chi }_{T}}(r,\theta ).
	\label{eqA17}
	\end{eqnarray}
	It can be seen that the following expression can be considered as an effective geometrical potential which can be regarded as a perturbation potential:
	\begin{eqnarray}
	{}^{'}{{V}^{'}}=\frac{{{\hbar }^{2}}}{2m}\alpha f(r)[\partial _{r}^{2}+\frac{1}{r}(1-\frac{2{{r}^{2}}}{{{b}^{2}}}){{\partial }_{r}}-\frac{{{r}^{2}}}{{{b}^{4}}}].
	\label{eqA18}
	\end{eqnarray}
	The corresponding Green's function for flat surface should satisfy
	\begin{eqnarray}
	[E+\frac{{{\hbar }^{2}}}{2m}(\partial _{r}^{2}+\frac{1}{r}{{\partial }_{r}}+\frac{1}{{{r}^{2}}}\partial _{\theta }^{2})]{{G}_{0}}(r,\theta ;{r}',{\theta }')=\frac{1}{\sqrt{-g}}\delta (r-{r}')\delta (\theta -{\theta }').
	\label{eqA19}
	\end{eqnarray}
	The solution of Eq. \eqref{eqA9} is: $G_{KG}^{0}(\vec{x},\vec{{x}'})=\frac{-i}{4}\frac{2m}{{{\hbar }^{2}}}H_{0}^{(1)}(\sqrt{\frac{2mE}{{{\hbar }^{2}}}}| \vec{x}-\vec{{x}'} |)$.          
	The Dyson expansion can be employed to obtain the Green's propagator in the presence of the perturbation potential up to the first order correction, as given in Eq. \eqref{eq19} .  
\end{widetext}
\section{The (2 + 1) Curved Dirac Equation in Polar Coordinates}
\label{B}
When the curvature effect in the Schr\"{o}dinger equation develops as a geometric potential formed from kinetic energy and the Costa term, we might expect a similar impact in the Dirac equation in curved space-time. The following modifications to the Dirac equation must be made to achieve the Dirac equation in curved space-time.
First, we have to convert the partial derivative into a covariant derivative; second, we have to consider the limiting potential; and third, we have to consider contribution of kinetic energy of the perpendicular degree of freedom $(m({{q}^{3}}){{v}_{F}}^{2})$. So, we can write Dirac equation as follows:
\begin{eqnarray}
D\Psi ({{q}^{i}})=[-i\hbar {{v}_{F}}{{\gamma }^{\mu }}({{\partial }_{\mu }}+{{\Omega }_{\mu }})+m({{q}^{3}}){{v}_{F}}^{2}\nonumber 
\\+{{V}_{\bot }}({{q}^{3}})]\Psi ({{q}^{i}})= 0,~~ i=0,1,2,3
\label{eqB1}
\end{eqnarray}

where ${{v}_{F}}$  is the Fermi velocity in a Dirac material such as graphene, $({{q}^{1}},{{q}^{2}})$ are the generalized coordinates of the surface, ${{q}^{3}}$ is normal coordinate and ${{D}_{a}}$ is covariant derivative standing for a (3+1)-dimensional space. ${{\gamma }^{\mu }}$ are curved gamma matrices that satisfy: $\left\{ {{\gamma }^{\mu }},{{\gamma }^{\nu }} \right\}=2{{g}^{\mu \nu }}$.
To obtain each of the terms in Eq. \eqref{eqB1}, we have to introduce some parameters. In this method, it is supposed that,
the effective mass term $m({{q}^{3}})$ has the following property:
\begin{eqnarray}
m(q^{3})=\begin{cases}
0~~~~~q^{3}=0,\\
m~~~~ q^{3} \ne 0
\label{eqB2}
\end{cases}.
\end{eqnarray}
The restricting potential $V_{\bot }(q^3)$ is what keeps the particle to surface.
\begin{eqnarray}
V_{\bot }(q^3)=\begin{cases}
0~~~~~q^{3}=0\\
\infty~~~~q^{3}\ne 0
\end{cases}.
\label{eqB3}
\end{eqnarray}
For a flat space-time, the line element is given by
$d{{s}^{2}}={{\eta}_{ab}}d{{x}^{a}}d{{x}^{b}};a,b=0,1,2,3$ and ${{\eta }_{ab}}=diag(1,-1,-1,-1)$.
For the curved space-time, regarding the Eq. \eqref{eqA1}, we obtain the metric of the manifold embedded in 3D space.
\begin{widetext}
\begin{eqnarray}
d{{s}^{2}}&=&{{G}_{{\mu }'{\nu }'}}d{{q}^{{{\mu }'}}}d{{q}^{{{\nu }'}}} \nonumber \\&=&{{G}_{00}}d{{q}^{0}}d{{q}^{0}}-{{\partial }_{\mu }}\vec{R}.{{\partial }_{\nu }}\vec{R}d{{q}^{\mu }}d{{q}^{\nu }}-{{\partial }_{\mu }}\vec{R}.{{\partial }_{3}}\vec{R}d{{q}^{\mu }}d{{q}^{3}}
	 -{{\partial }_{3}}\vec{R}.{{\partial }_{\mu }}\vec{R}d{{q}^{3}}d{{q}^{\nu }}-{{\partial }_{3}}\vec{R}.{{\partial
	}_{3}}\vec{R}~d{{q}^{3}}d{{q}^{3}}.
\label{eqB4}
\end{eqnarray}
\end{widetext}
In that follows, we shall find a relationship in 3D space between the spatial components of the two-coordinate, surface, and Euclidean metrics.
\begin{eqnarray}
d{{s}^{2}}={{G}_{00}}d{{q}^{0}}d{{q}^{0}}-{{G}_{\mu \nu }}d{{q}^{\mu }}d{{q}^{\nu }},~\mu,\nu =1,2,3.	
\label{eqB5}
\end{eqnarray}
Using Eq.\eqref{eqA1}, it can be obtained that
\begin{widetext}
	\begin{eqnarray}
	d{{s}^{2}}&=&{{G}_{00}}d{{q}^{0}}d{{q}^{0}}-[(\delta _{\mu }^{\lambda }-{{q}^{3}}\alpha _{\mu }^{\lambda }){{\partial }_{\lambda }}\vec{r}.(\delta _{\nu }^{\beta }-{{q}^{3}}\alpha _{\nu }^{\beta }){{\partial }_{\beta }}\vec{r}]d{{q}^{\mu }}d{{q}^{\nu }}-d{{q}^{3}}d{{q}^{3}} 
	\nonumber \\&=&-[{{g}_{\mu \nu }}-{{q}^{3}}(\alpha _{\mu }^{\lambda }{{g}_{\lambda \nu }}+\alpha _{\nu }^{\beta }{{g}_{\beta \mu }})+{{({{q}^{3}})}^{2}}\alpha _{\mu }^{\lambda }{{g}_{\lambda \beta }}\alpha _{\nu }^{\beta }]d{{q}^{\mu }}d{{q}^{\nu }}-d{{q}^{3}}d{{q}^{3}},  
	\label{eqB6}
	\end{eqnarray}
\end{widetext}
and for brevity's sake, the following relation is defined as 
\begin{eqnarray}
{{\gamma }_{\mu \nu }}={{g}_{\mu \nu }}-{{q}^{3}}(\alpha _{\mu }^{\lambda }{{g}_{\lambda \nu }}+\alpha _{\nu }^{\beta }{{g}_{\beta \mu }})+{{({{q}^{3}})}^{2}}\alpha _{\mu }^{\lambda }{{g}_{\lambda \beta }}\alpha _{\nu }^{\beta }~~~~~~
\label{eqB7}
\end{eqnarray}
and finally we get 
\begin{eqnarray}
d{{s}^{2}}=-{{\gamma }_{\mu \nu }}d{{q}^{\mu }}d{{q}^{\nu }}-d{{q}^{3}}d{{q}^{3}}
\label{eqB8}.
\end{eqnarray}
For the curved space-time, we can write
\begin{eqnarray}
d{{s}^{2}}&=&{{G}_{{\mu }'{\nu }'}}d{{q}^{{{\mu }'}}}d{{q}^{{{\nu }'}}}\nonumber \\
&=&-{{\gamma }_{\mu \nu }}d{{q}^{\mu }}d{{q}^{\nu }}-d{{q}^{3}}d{{q}^{3}}
\label{eqB9}
\end{eqnarray}
where ${\mu }',{\nu }'$ run over $(0,...,3)$ also $\mu ,\nu = (0,...,2)$. 
Meanwhile, for the flat space-time we have
\begin{eqnarray}
d{{s}^{2}}&=&{{\eta }_{{a}'{b}'}}d{{x}^{{{a}'}}}d{{x}^{{{b}'}}}\nonumber \\&=&{{\eta }_{ab}}d{{x}^{a}}d{{x}^{b}}-d{{x}^{3}}d{{x}^{3}}, \end{eqnarray}        
where ${a }',{b }'$ run over $(0,...,3)$ also $a ,b = (0,...,2)$, \\
in which :
${{e}_{{{\mu }'}}}=\frac{\partial }{\partial {{q}^{{{\mu }'}}}}\equiv {{\partial }_{{{\mu }'}}}$, ${{e}_{{{a}'}}}=\frac{\partial }{\partial {{x}^{{{a}'}}}}\equiv {{\partial }_{{{a}'}}}$.\\
Conversion of the curved coordinate to the flat space is given by
\begin{eqnarray}
{{e}_{{{a}'}}}=\frac{\partial }{\partial {{x}^{{{a}'}}}}=\frac{\partial {{q}^{{{\mu }'}}}}{\partial {{x}^{{{a}'}}}}\frac{\partial }{\partial {{q}^{{{\mu }'}}}}=e_{{{a}'}}{}^{{{\mu }'}}{{e}_{{{\mu }'}}}. 
\label{eqB11}                 
\end{eqnarray}
\\
The coefficients $e_{{{a}'}}{}^{{{\mu }'}}$ are known as vierbein coefficients, and we denote $e_{3{a}'}{}^{{{\mu }'}}$ and $e_{2{a}'}{}^{{{\mu }'}}$,  for the coefficients of the 3D space, and 2D curved surface, respectively. The following are the conversion equations between the two metrics, global coordinate, and local coordinate:
\begin{eqnarray}
{{\eta }_{{a}'{b}'}}&=&{{\partial }_{{{a}'}}}\vec{R}.{{\partial }_{{{b}'}}}\vec{R}\nonumber \\
&=&\frac{\partial {{q}^{{{\mu }'}}}}{\partial {{x}^{{{a}'}}}}\frac{\partial {{q}^{{{\nu }'}}}}{\partial {{x}^{{{b}'}}}}{{\partial }_{{{\mu }'}}}\vec{R}.{{\partial }_{{{\nu }'}}}\vec{R}\nonumber\\ 
&=&e_{3{a}'}{}^{{{\mu }'}}.e_{3{b}'}{}^{{{\nu }'}}{{G}_{{\mu }'{\nu }'}}. 
\label{eqB12}
\end{eqnarray}
The general relation between metrics of flat space-time and curved space-time is given in Eq. \eqref{eqB12}.
Meanwhile, one can write
\begin{eqnarray}
\label{eqB13}
\eta_{ab}&=&{\partial_{a}}\vec{R}.{\partial }_{b}\vec{R}\nonumber\\
&=&\frac{\partial {{q}^{\mu }}}{\partial {{x}^{a}}}\frac{\partial {{q}^{\nu }}}{\partial {{x}^{b}}}{{\partial }_{\mu }}\vec{R}.{{\partial }_{\nu }}\vec{R} \\ 
&=&e_{3a}{}^{\mu }.e_{3b}{}^{\nu }{{\gamma }_{\mu \nu }}.
\nonumber 
\end{eqnarray}
Equation\eqref{eqB13} is a subset of the previous one, but it lacks any contributions from the vertical component on the surface. On the other hand,
\begin{eqnarray}
{{\eta }_{ab}}&=&{{\partial }_{a}}\vec{r}.{{\partial }_{b}}\vec{r}\nonumber \\&=&\frac{\partial {{q}^{\mu }}}{\partial {{x}^{a}}}\frac{\partial {{q}^{\nu }}}{\partial {{x}^{b}}}{{\partial }_{\mu }}\vec{r}.{{\partial }_{\nu }}\vec{r} \\ 
&=&e_{2a}{}^{\mu }.e_{2b}{}^{\nu }{{g}_{\mu \nu }}. \nonumber 
\label{eqB14}
\end{eqnarray}
\\
The metric of the 2D area of a curved surface is defined by Eq. (B14), where we have
\begin{eqnarray}
{{e}_{3a}}^{\mu }{{e}^{b}}_{3\mu }=\delta _{a}^{b} \nonumber \\ {{e}_{3a}}^{\mu }{{e}^{a}}_{3\nu }=\delta _{\nu }^{\mu },
\label{eqB15}
\end{eqnarray}
it can easily be obtained by Eqs. \eqref{eqB13} and \eqref{eqB14} that:
\begin{eqnarray}
{{e}^{a}}_{3\mu }={{e}^{a}}_{2\mu }+{{q}^{3}}\alpha _{\mu }^{\nu }{{e}^{a}}_{2\nu }.
\label{eqB16}
\end{eqnarray}
We treat ${{q}^{3}}$ as a perturbation parameter because we must reduce the normal dimension, ${{q}^{3}}$, to limit the charge carriers remain on the surface. Let's utilize $\varepsilon $ as a perturbation parameter and rescale the normal coordinates by ${{q}^{3}}\to \varepsilon {{q}^{3}}$:
\begin{eqnarray}
{{e}^{a}}_{3\mu }={{e}^{a}}_{2\mu }+\varepsilon {{q}^{3}}\alpha _{\mu }^{\nu }{{e}^{a}}_{2\nu }\nonumber \\
{{e}^{\mu}}_{3a }={{e}^{\mu}}_{2a }-\varepsilon{{q}^{3}}\alpha _{\mu }^{\nu }{{e}^{a}}_{2\nu }+0({{\varepsilon }^{2}}). 
\label{eqB17}
\end{eqnarray}
To calculate spin connection, the following nonzero 3D Christoffel symbols must be known
\begin{eqnarray}
\Gamma _{\mu \lambda }^{\nu }=\frac{1}{2}{{G}^{\nu k}}({{\partial }_{\mu }}{{G}_{\lambda k}}+{{\partial }_{\lambda }}{{G}_{\mu k}}-{{\partial }_{k}}{{G}_{\mu \lambda }})
\label{eqB18}
\end{eqnarray}
\begin{eqnarray}
\Gamma _{\mu \lambda }^{3}=\frac{1}{2}{{\eta }^{33}}(2{{\alpha }_{\mu \lambda }})=-{{\alpha }_{\mu \lambda }}~~~~~~~~~~~~~~
\label{eqB19} 
\end{eqnarray}
\begin{eqnarray}
\Gamma _{3\lambda }^{\nu }&=&\frac{1}{2}{{G}^{\nu k}}({{\partial }_{3}}{{G}_{\lambda k}}+{{\partial }_{\lambda }}{{G}_{3k}}-{{\partial }_{k}}{{G}_{3\nu }}) \nonumber 
\\ 
& =&\frac{1}{2}{{G}^{\nu k}}(-2{{\alpha }_{\lambda k}})\nonumber \\&=&-\alpha _{\lambda }^{\nu }.
\label{eqB20}
\end{eqnarray}
The following is derived by replacing Eqs.\eqref{eqB19} and \eqref{eqB20} into \eqref{eq26}:
\begin{widetext}
	\begin{eqnarray}
	{{\Omega }_{\mu }}&=&\frac{1}{4}{{\gamma }^{a}}{{\gamma }^{b}}{{\eta }_{ac}}{e^c}_{3\nu }
	({e_{3b}}^{\lambda }\Gamma _{\mu \lambda }^{\nu }+{{\partial }_{\mu }}{{e_{3b}}}^{\nu })-\frac{1}{4}{{\gamma }^{a}}{{\gamma }^{3}}{{e}_{2a\nu }}\alpha _{\mu }^{\nu }+\frac{1}{4}{{\gamma }^{3}}{{\gamma }^{a}}{{g}_{\lambda \nu }}{e_{2a}}^{\lambda }\alpha _{\mu }^{\nu }+O(\varepsilon )\nonumber \\&=&\frac{1}{4}{{\gamma }^{a}}{{\gamma }^{b}}{{\eta }_{ac}}{e^{c}}_{3\nu }({e_{3b}}^{\lambda }\Gamma _{\mu \lambda }^{\nu }+{{\partial }_{\mu }}{e_{3b}}^{\nu })-\frac{1}{2}{{\gamma }^{a}}{{\gamma }^{3}}{{e}_{2a\nu }}\alpha _{\mu }^{\nu }+O(\varepsilon )\ ,\mu ,\lambda ,\nu =1,2.
	\label{eqB21}
	\end{eqnarray}
	Similarly,  by replacing Eq.\eqref{eqB17} into \eqref{eqB21}:
	\begin{eqnarray}
	\begin{aligned}
	{{\Omega }_{3}}&=\frac{1}{4}{{\gamma }^{a}}{{\gamma }^{b}}{{\eta }_{ac}}{e^{c}}_{3\nu }({e_{3b}}^{\lambda }\Gamma _{3\lambda }^{\nu }+{{\partial }_{3}}{e_{3b}}^{\nu })\nonumber \\&=\frac{1}{4}{{\gamma }^{a}}{{\gamma }^{b}}{{\eta }_{ac}}({e^{c}}_{2\nu }+\varepsilon {{q}^{3}}\alpha _{\nu }^{{{\beta }_{1}}}{e^{c}}_{2{{\beta }_{1}}})[({e_{2b}}^{\lambda }-\varepsilon {{q}^{3}}\alpha _{{{\beta }_{3}}}^{\lambda }{e_{2b}}^{{{\beta }_{3}}})\Gamma _{3\lambda }^{\nu }+{{\partial }_{3}}({e_{2b}}^{\nu }-\varepsilon {{q}^{3}}\alpha _{{{\beta }_{2}}}^{\nu }{e_{2b}}^{{{\beta }_{2}}})] \nonumber ~~~~~~~~\\
	& =\frac{1}{4}{{\gamma }^{a}}{{\gamma }^{b}}{{\eta }_{ac}}({-e^{c}}_{2\nu }\alpha _{{{\beta }_{2}}}^{\nu }{e_{2b}}^{{{\beta }_{2}}}+e_{2\nu }^{c}\Gamma _{3\lambda }^{\nu }{e_{2b}}^{\lambda })=\frac{1}{4}{{\gamma }^{a}}{{\gamma }^{b}}{{\eta }_{ac}}({-e^{c}}_{2\nu }\alpha _{{{\beta }_{2}}}^{\nu }{e_{2b}}^{{{\beta }_{2}}}+{e^{c}}_{2\nu }\alpha _{\lambda }^{\nu }{e_{2b}}^{\lambda }) ~~~~~~~~\\ &=0,~~~~~~~~~~~~~~~~~~~~~~~~~~~~~~~~~~~~~~~~~~~~~~~~~~~~~~~~~~~~~~~~~~~~~~~~~~~~~~~~~~~~~~~~~~~~~~~~~~~~\nonumber
	\end{aligned}
	\end{eqnarray}\\
	also
	${{\Omega }_{0}}=0.$\\
To provide an effective dynamical equation on the surface, $\Psi(t,{{q}^{{1}}},{{q}^{2}},{{q}^{3}})$ as well as the Dirac operator $D$ should be converted. The Dirac equation with a surface wave function can be found by performing the following substitutions \cite{6,14,15}:\\

	$\chi (t,{{q}^{{1}}},{{q}^{2}},{{q}^{3}})={{(\frac{G}{g})}^{\frac{1}{4}}}\Psi (t,{{q}^{{1}}},{{q}^{2}},{{q}^{3}}) \\ ~~~~~~~~~~~~~~~~~~~~~~=\sqrt{\Lambda }\Psi (t,{{q}^{{1}}},{{q}^{2}},{{q}^{3}})\\
	~~~~~~~~~~~~~~~~~~~~\Lambda =1+\varepsilon {{q}^{3}}Tr\alpha +{{(\varepsilon {{q}^{3}})}^{2}}Det(\alpha )\\	
	~~~~~~~~~~~~~~~~~~~D\equiv {{(\frac{G}{g})}^{\frac{1}{4}}}D{{(\frac{g}{G})}^{\frac{1}{4}}}.$\\
	
	After some algebra, we have found the following relation:
	\begin{eqnarray}
	D\chi ({{q}^{i}})\equiv {{(\frac{G}{g})}^{\frac{1}{4}}}D{{(\frac{g}{G})}^{\frac{1}{4}}}\chi ({{q}^{i}})
	=-i\hbar {{v}_{F}}{{\gamma }^{\mu }}({{\partial }_{\mu }}+{{\Omega }_{\mu }})\chi ({{q}^{i}})+\frac{1}{2}i\hbar {{v}_{F}}{{\gamma }^{3}}tr\gamma \chi ({{q}^{i}})+{{(\frac{G}{g})}^{\frac{1}{4}}}i\hbar {{v}_{F}}{{\gamma }^{3}}\frac{{{\partial }_{3}}}{\varepsilon }[{{(\frac{g}{G})}^{\frac{1}{4}}}\chi ({{q}^{i}})]\nonumber \\+  
	m(\varepsilon {{q}^{3}}){{v}_{F}}^{2}\chi ({{q}^{i}})+{{V}_{\bot }}(\varepsilon {{q}^{3}})\chi ({{q}^{i}})=0. 
	\label{eqB22}
	\end{eqnarray}
	The third term is now provided by
\begin{eqnarray}
	{{(\frac{G}{g})}^{\frac{1}{4}}}\frac{{{\partial }_{3}}}{\varepsilon }[{{(\frac{g}{G})}^{\frac{1}{4}}}\chi ({{q}^{i}})]&=&\frac{1}{\varepsilon }{{\partial }_{3}}\chi ({{q}^{i}})-\chi ({{q}^{i}})\frac{1}{2}\frac{tr(\alpha) }{(1+\varepsilon {{q}^{3}}tr\alpha +{{(\varepsilon {{q}^{3}})}^{2}}Det(\alpha ))}.\nonumber
\end{eqnarray}
	Using the  $\varepsilon \ll 1$ approximation, we get
	\begin{eqnarray}
	[\frac{{{\partial }_{3}}}{\varepsilon }-\frac{tr(\alpha) }{2}(1-\varepsilon {{q}^{3}}tr(\alpha) +O({{\varepsilon }^{2}}))]\chi ({{q}^{i}})=[\frac{{{\partial }_{3}}}{\varepsilon }-\frac{tr(\alpha) }{2}+O(\varepsilon )]\chi ({{q}^{i}}).
	\label{eqB23}
	\end{eqnarray}
	The following results can simply be obtained by inserting \eqref {eqB1} into the Dirac equation:
	\begin{eqnarray}
	-i\hbar {{v}_{F}}e_{f}^{\mu }{{\gamma }^{f}}({{\partial }_{\mu }}+{{\Omega }_{\mu }})\chi ({{q}^{i}})-\frac{1}{2}i\hbar {{v}_{F}}{{\gamma }^{3}}tr(\gamma) \chi ({{q}^{i}})-i\hbar {{v}_{F}}{{\gamma }^{3}}[{{\partial }_{3}}-\frac{tr(\alpha) }{2}+O(\varepsilon )]\chi ({{q}^{i}})&+&m(\varepsilon {{q}^{3}}){{v}_{F}}^{2}\chi ({{q}^{i}})+{{V}_{\bot }}(\varepsilon {{q}^{3}})\chi ({{q}^{i}})\nonumber
	\\&=&0.
	\label{eqB24}
	\end{eqnarray}
	Using the following relation
	\begin{eqnarray}
	{{\gamma }^{\mu }}{{\gamma }^{a}}{{\gamma }^{3}}{{\eta }_{ac}}\alpha _{\mu }^{\nu }e_{2\nu }^{c}={{\gamma }^{3}}tr(\alpha ),
	\label{eqB25}
	\end{eqnarray}\nonumber
	we can obtain
	\begin{eqnarray}
	-i\hbar {{v}_{F}}{{\gamma }^{\mu }}({{\partial }_{\mu }}+{{\Omega }_{\mu }})\chi ({{q}^{i}})-i\hbar {{v}_{F}}{{\gamma }^{3}}\frac{{{\partial }_{3}}}{\varepsilon }\chi ({{q}^{i}})+m(\varepsilon {{q}^{3}}){{v}_{F}}^{2}\chi ({{q}^{i}})+{{V}_{\bot }}(\varepsilon {{q}^{3}})\chi ({{q}^{i}})=0,~~~\mu =1,2.
	\label{eqB26}
	\end{eqnarray}
	This demonstrates that, no Costa potential is created for Dirac carriers up to the first order. Separation of variables is used to solve the Dirac equation
	\begin{eqnarray}
	\label{eqB27}
	\chi (t,{{q}^{{1}}},{{q}^{2}},{{q}^{3}})={{\chi }_{T}}({{q}^{{1}}},{{q}^{2}},t).{{\chi }_{N}}({{q}^{3}},t).
	\end{eqnarray}
Inserting Eq.\eqref{eqB27} into \eqref{eqB26} results in
	\begin{eqnarray}
	\label{eqB}
	\frac{1}{{{\chi }_{T}}(t,{{q}^{{1}}},{{q}^{2}})}\{-i\hbar {{v}_{F}}[{{\gamma }^{\mu }}({{\partial }_{\mu }}+{{\Omega }_{\mu }})+{{\gamma }^{0}}{{\partial }_{t}}]{{\chi }_{T}}(t,{{q}^{{1}}},{{q}^{2}})\}+\frac{1}{{{\chi }_{N}}(t,{{q}^{3}})}[-i\hbar {{v}_{F}}({{\gamma }^{3}}\frac{{{\partial }_{3}}}{\varepsilon }+{{\gamma }^{0}}{{\partial }_{t}})  
	+m(\varepsilon {{q}^{3}}){{v}_{F}}^{2}+\nonumber\\{{V}_{\bot }}(\varepsilon {{q}^{3}})]{{\chi }_{N}}(t,{{q}^{3}})=0. 
	\end{eqnarray}
	Because $ {{\chi }_{T}}({{q}^{{1}}},{{q}^{2}},t)$ and ${{\chi }_{N}}({{q}^{3}},t)$ are independent of each other, we get the following equations
\begin{eqnarray}
	\label{eqB28}
	\frac{1}{{{\chi }_{T}}(t,{{q}^{{1}}},{{q}^{2}})}\{-i\hbar {{v}_{F}}[{{\gamma }^{\mu }}({{\partial }_{\mu }}+{{\Omega }_{\mu }})+{{\gamma }^{0}}{{\partial }_{t}}]{{\chi }_{T}}(t,{{q}^{{1}}},{{q}^{2}})\}=0,~~~~~~\mu =1,2.
\end{eqnarray}
	
We get the following equation by plugging the values of $\Omega$ into the above equation
	\begin{eqnarray}
	\frac{1}{{{\chi }_{N}}(t,{{q}^{3}})}[-i\hbar {{v}_{F}}({{\gamma }^{3}}\frac{{{\partial }_{3}}}{\varepsilon }+{{\gamma }^{0}}{{\partial }_{t}})+m(\varepsilon {{q}^{3}}){{v}_{F}}^{2}+{{V}_{\bot }}(\varepsilon {{q}^{3}})]{{\chi }_{N}}(t,{{q}^{3}})=0.
	\label{eqB29}
	\end{eqnarray}
	The first Eq. \eqref{eqB28} can be generalized:
	\begin{eqnarray}
	-i\hbar {{v}_{F}}{{\gamma }^{\mu }}({{\partial }_{\mu }}+{{\Omega }_{\mu }}){{\chi }_{T}}(t,{{q}^{_{1}}},{{q}^{2}})=0, ~~\mu =0,1,2.
	\label{eqB30}
	\end{eqnarray}
	This equation is exactly the Eq. \eqref{eq28} that we have employed in this paper. It does not contain the so called geometric potential, which, as shown before, can be neglected at least up to first order. Meanwhile, by expanding the covariant derivative, an effective potential (non-geometric potential) is obtained that shows a direct relation to the curvature of the surface.
	In the following, if we consider the particle dynamics along the normal direction, one can write
	$m(\varepsilon {{q}^{3}})\approx 0 $, and ${{V}_{\bot }}(\varepsilon {{q}^{3}})\approx \frac{{{V}_{\bot }}({{q}^{3}})}{\varepsilon }$, then \[[-i\hbar {{v}_{F}}({{\gamma }^{3}}\frac{{{\partial }_{3}}}{\varepsilon }+{{\gamma }^{0}}{{\partial }_{t}})+\frac{{{V}_{\bot }}({{q}^{3}})}{\varepsilon }]{{\chi }_{N}}(t,{{q}^{3}})=0.\]
	where it was assumed that \[{{\chi }_{N}}(t,{{q}^{3}})={{e}^{i{{\omega }_{n}}t}}{{\chi }_{N}}({{q}^{3}})\],
	then the following relation can be obtained:  
	\[-i\hbar {{v}_{F}}({{\gamma }^{3}}{{\partial }_{3}}+{{V}_{\bot }}({{q}^{3}})){{\chi }_{N}}({{q}^{3}})+\varepsilon (\hbar {{v}_{F}}\gamma {{\omega }_{n}}){{\chi }_{N}}({{q}^{3}})=0\].
	In the limit of  $\varepsilon \to 0$, then
	$[-i\hbar {{v}_{F}}{{\gamma }^{3}}{{\partial }_{3}}+{{V}_{\bot }}({{q}^{3}})]{{\chi }_{N}}({{q}^{3}})=0$ and using to the normalization condition for the vertical component, we can obtain ${{\chi }_{N}}({{q}^{3}})={{e}^{\frac{i{{\gamma }^{-3}}}{\hbar {{v}_{F}}}\int_{0}^{{{q}^{3}}}{{{V}_{\bot }}({{q}^{3}})d}{{q}^{3}}}}$.
	This demonstrates that ${{\chi }_{N}}(t,{{q}^{3}})$ cannot change the real space probability of the particles or local density of the sample since one can realize that: 
	${{\left| \chi ({{q}^{i}}) \right|}^{2}}={{\left| {{\chi }_{t}}(t,{{q}^{1}},{{q}^{2}}) \right|}^{2}}{{\left| {{\chi }_{N}}(t,{{q}^{3}}) \right|}^{2}}={{\left| {{\chi }_{t}}(t,{{q}^{1}},{{q}^{2}}) \right|}^{2}}$.\\
\end{widetext}
The details of the calculations, which have been mentioned in Sec. IV, are presented below. First, the outcomes of the first choice, Eq. (\ref{eq30}), for a flat surface and then for a curved surface are obtained.  After these calculations, some geometric parameters of the first choice of verbiens are obtained for both flat and curved surfaces.
In the flat case, the non-zero Christoffel coefficients in polar coordinates are
\begin{eqnarray} 
\Gamma _{\theta \theta }^{r}=-r~~~~~~~~~~~~~~~~~~~~\nonumber\\
\Gamma _{r\theta }^{\theta }=\Gamma _{\theta r}^{\theta }=\frac{1}{r}~~~~~~~~~~~~~
\label{eqB31}
\end{eqnarray}

The geometric factors of the first choice are:
\begin{eqnarray} 
{e^{a}}_{\mu }&=&\begin{pmatrix}
1 & 0  \\
0 & r  \\
\end{pmatrix},
\nonumber\\ 
{e_{a}}^{\mu }&=&\begin{pmatrix}
1 & 0  \\
0 & \frac{1}{r}  \\
\end{pmatrix}.  
\label{eqB32}
\end{eqnarray}
It can also be obtained for the non-zero $\omega$ parameter,
\begin{eqnarray} 
{{\omega }_{\theta 12 }}=-{{\omega }_{\theta 21}}=1,~~~~~~~~~~~~~~~~~~
\label{eqB33}
\end{eqnarray}
and, eventually, the flat Dirac equation is obtained as follows:
\begin{eqnarray} 
i\hbar {{v}_{f}}({{\gamma }^{0}}\frac{{{\partial }_{t}}}{{{v}_{f}}}+{{\gamma }^{1}}({{\partial }_{r}}+\frac{1}{2r})+\frac{{{\gamma }^{2}}}{r}{{\partial }_{\theta }})\psi (t,r,\theta )=0.
\label{eqB34}~~~~
\end{eqnarray}
The term $\frac{{{\gamma }^{1}}}{2r}$ is originated from the spinor nature of the Dirac wave function. This term also appears as a result of the graphene honeycomb structure which contains two $A$ and $B$ sublattices. The Dirac Hamiltonian of the flat space in polar coordinates can be written as
\begin{eqnarray} 
H_{flat}=\hbar {{\nu }_{f}}\begin{pmatrix} 
0 & {{\partial }_{r}}-i\frac{{{\partial }_{\theta }}}{r}+\frac{1}{2r}  \\
{{\partial }_{r}}+i\frac{{{\partial }_{\theta }}}{r}+\frac{1}{2r} & 0  \\
\end{pmatrix}. 
\label{eqB35}
\end{eqnarray}
The above parameters are now assessed for the first case in a curved surface. In the polar coordinate system, the curved state metric yields non-zero Christoffel coefficients, vierbeins, the non-zero spin connection, and the spinoral affine connection as follows:
\begin{eqnarray}
\label{eqB36}
\Gamma _{rr}^{r}&=&\frac{1}{2}\frac{\alpha {f}'(r)}{1+\alpha f(r)}\nonumber \\
\Gamma _{\theta \theta }^{r}&=&\frac{-r}{1+\alpha f(r)}
\\   
\Gamma _{r\theta }^{\theta }&=&\Gamma _{\theta r}^{\theta }=\frac{1}{r},\nonumber  
\end{eqnarray}
and the fist case vierbiens in a curved surface are:
\begin{eqnarray}
{e^{a}}_{\mu }=
\begin{pmatrix}
{{(1+\alpha f(r))}^{\frac{1}{2}}} & 0  \\
0 & r  \\
\end{pmatrix}
\nonumber\\
{e_{a}}^{\mu } =\begin{pmatrix}
\frac{1}{{{(1+\alpha f(r))}^{\frac{1}{2}}}} & 0  \\
0 & \frac{1}{r}  \\
\end{pmatrix}.
\label{eqB37}
\end{eqnarray}
Meanwhile, the non-zero $\omega$ in a curved surface are
\begin{eqnarray}
 {{\omega }_{\theta 12 }}=-{{\omega }_{\theta 21}}={{(1+\alpha f(r))}^{-\frac{1}{2}}}.
\label{eqB38}
\end{eqnarray}\\
By substituting Eqs. \eqref{eqB35}-\eqref{eqB37} into Eq. \eqref{eqB30}, we get the following modified Dirac equation:\\
\\
$i\hbar {{v}_{f}}({{\gamma }^{0}}\frac{{{\partial }_{t}}}{{{v}_{f}}}+\frac{{{\gamma }^{1}}}{{{(1+\alpha f(r))}^{\frac{1}{2}}}}({{\partial }_{r}}+\frac{1}{2r})+ \frac{{{\gamma }^{2}}}{r}{{\partial }_{\theta }})\psi (t,r,\theta )=0.$ 
\\
\\
The corresponding Hamiltonian is:
\begin{eqnarray}
H{}_{curved}=\hbar {{\nu }_{f}}\begin{pmatrix}
0 & \frac{({{\partial }_{r}}+\frac{1}{2r})}{{{(1+\alpha f(r))}^{\frac{1}{2}}}}-i\frac{{{\partial }_{\theta }}}{r}  \\
\frac{({{\partial }_{r}}+\frac{1}{2r})}{{{(1+\alpha f(r))}^{\frac{1}{2}}}}+i\frac{{{\partial }_{\theta }}}{r} & 0  \\
\end{pmatrix}. \nonumber\\
\label{eqB39}
\end{eqnarray}	
Next, we have to obtain non-zero key  geometric parameters of the second case. These verbiens were used in rest of the calculations where the RKKY interaction and LDOS are obtained.
In the flat case using Eqs. \eqref{eqB29} and \eqref{eqB30}, it can be realized that
\begin{eqnarray} 
{{\omega }_{\theta 22 }}&=&-\sin 2\theta. ~~~~~~~~~~~~~~~~~~~~    
\label{eqB40}
\end{eqnarray}
In the curved case, using Eq. \eqref{eqB37}, non-zero components of $\omega$ are obtained as
\begin{eqnarray}
{{\omega }_{r11}}&=&{{\cos }^{2}}\theta \frac{1}{2}\frac{\alpha {f}'(r)}{1+\alpha f(r)}{{[1-(1+\alpha f(r))]}^{\frac{1}{2}}},\nonumber \\ {{\omega }_{r12 }}&=&{{\omega }_{r21}}=\cos \theta \sin \theta \frac{1}{2}\frac{\alpha {f}'(r)}{1+\alpha f(r)}{{[1-(1+\alpha f(r))]}^{\frac{1}{2}}},\nonumber \\
{{\omega }_{r12 }}&=&{{\sin }^{2}}\theta \frac{1}{2}\frac{\alpha {f}'(r)}{1+\alpha f(r)}{{[1-(1+\alpha f(r))]}^{\frac{1}{2}}},\\ {{\omega }_{\theta 12 }}&=&-{{\omega }_{\theta 21 }}=1-{{(1+\alpha f(r))}^{-\frac{1}{2}}}.\nonumber 
\label{eqB41}
\end{eqnarray}

\section{Landau energy levels in graphene}
\label{C}
In a magnetic field, the motion of relativistic charges in graphene is quantized. Due to the parabolic dispersion law of free electrons, Landau quantization yields equidistant energy levels in a typical non-relativistic electron gas. The electrons in graphene have a relativistic dispersion law, which changes the the position of  Landau levels significantly. We use substitution $\vec{p}\to \vec{p}+e\vec{A}$ for free electrons. As a result, we will obtain:
\begin{eqnarray}
H={{v}_{f}}\begin{pmatrix}
0 & \vec{\sigma }.(\vec{p}+e\vec{A})  \\
-{{{\vec{\sigma }}}^{*}}.(\vec{p}+e\vec{A}) & 0  \\
\end{pmatrix},
\label{eqC1}
\end{eqnarray}
 which is a the four-component wave function 
\begin{eqnarray}
\Psi =\begin{pmatrix}
{{\varphi }_{{{k}_{A}}}}  \\
{{\varphi }_{{{k}_{B}}}}  \\
\end{pmatrix},\nonumber
\end{eqnarray} corresponding to the Hamiltonian eigenstate shown in Eq. \eqref{eqC1}. As a result, ${{\varphi }_{{{k}_{A}}}}$ and ${{\varphi }_{{{k}_{B}}}}$ are the wave functions for $K$ valley and $A$, $B$ sublattices, respectively. We get the following equations by solving the eigenvalue equation $H\Psi =E\Psi $:
\begin{widetext}
	\begin{eqnarray}
	{{v}_{f}}\begin{pmatrix}
	0 & ({{p}_{x}}+e{{A}_{x}})-i{{p}_{y}}  \\
	({{p}_{x}}+e{{A}_{x}})+i{{p}_{y}} & 0  \\
	\end{pmatrix}\begin{pmatrix}
	{{\varphi }_{{{k}_{A}}}}  \\
	{{\varphi }_{{{k}_{B}}}}  \\
	\end{pmatrix}=E\begin{pmatrix}
	{{\varphi }_{{{k}_{A}}}}  \\
	{{\varphi }_{{{k}_{B}}}}  \\
	\end{pmatrix}.	
	\label{eqC2}
	\end{eqnarray} 
\end{widetext}
In the following
\begin{eqnarray}
\begin{aligned}
& [({{p}_{x}}+e{{A}_{x}})-i{{p}_{y}}]{{\varphi }_{{{k}_{B}}}}=\frac{E}{{{v}_{f}}}{{\varphi }_{{{k}_{A}}}}, \\ 
& [({{p}_{x}}+e{{A}_{x}})+i{{p}_{y}}]{{\varphi }_{{{k}_{A}}}}=\frac{E}{{{v}_{f}}}{{\varphi }_{{{k}_{B}}}}. 
\end{aligned}
\label{eqC3}
\end{eqnarray}
We may get analogous equations for the ${K}'$ valley wave functions. 
By substituting and combining relations represented in Eq. \eqref{eqC3}:
\begin{eqnarray}
\begin{aligned}
& [({{p}_{x}}+e{{A}_{x}})+i{{p}_{y}}][({{p}_{x}}+e{{A}_{x}})-i{{p}_{y}}]{{\varphi }_{{{k}_{B}}}}={{(\frac{E}{{{v}_{f}}})}^{2}}{{\varphi }_{{{k}_{B}}}}, \\ 
& [({{p}_{x}}+e{{A}_{x}})-i{{p}_{y}}][({{p}_{x}}+e{{A}_{x}})+i{{p}_{y}}]{{\varphi }_{{{k}_{A}}}}={{(\frac{E}{{{v}_{f}}})}^{2}}{{\varphi }_{{{k}_{A}}}},
\label{eqC4}
\end{aligned}
\end{eqnarray}
then, we can derive the equations using the fact that\\
$[{{p}_{y}},{{p}_{x}}-eBy]=i\hbar eB$:
\begin{eqnarray}
\{{{({{p}_{x}}-eBy)}^{2}}+i[{{p}_{y}},{{p}_{x}}-eBy]+{{p}_{y}}^{2}\}{{\varphi }_{{{k}_{B}}}}={{(\frac{E}{{{v}_{f}}})}^{2}}{{\varphi }_{{{k}_{B}}}};~~~~
\label{eqC5}
\end{eqnarray}
\begin{eqnarray}
\{{{({{p}_{x}}-eBy)}^{2}}-i[{{p}_{y}},{{p}_{x}}-eBy]+{{p}_{y}}^{2}\}{{\varphi }_{kA}}={{(\frac{E}{{{v}_{f}}})}^{2}}{{\varphi }_{{{k}_{A}}}}~~~~
\label{eqC6}
\end{eqnarray}
If we define:
\begin{eqnarray}
{{y}_{0}}=\frac{\hbar {{k}_{x}}}{eB}.
\label{eqC7},
\end{eqnarray}
then we get:
\begin{eqnarray}
{{(eB)}^{2}}{{(y-{{y}_{0}})}^{2}}+{{p}_{y}}^{2}{{\varphi }_{{{k}_{B}}}}=[{{(\frac{E}{{{v}_{f}}})}^{2}}+eB\hbar ]{{\varphi }_{{{k}_{B}}}}.
\label{eqC8}
\end{eqnarray}  
In the presence of a magnetic field, the left hand side of Eq. \eqref{eqC7} is an oscillator equation with well-known definite solutions, allowing the energy of this system to be retrieved.   
For  ${{\varphi }_{{{k}_{B}}}}:  (n+\frac{1}{2})2e\hbar B=[{{(\frac{E}{{{v}_{f}}})}^{2}}+e\hbar B], $\\
we have
\begin{eqnarray}
&{{E}_{n}}=\pm {{v}_{f}}\sqrt{n2eB\hbar },n=0,1,2,... ,
\label{eqC9}
\end{eqnarray}   
and for  ${{\varphi }_{{{k}_{A}}}}:(n+\frac{1}{2})2e\hbar B=[{{(\frac{E}{{{v}_{f}}})}^{2}}-e\hbar B], $  \\                                                                        
we get
\begin{eqnarray}
{{E}_{n}}=\pm {{v}_{f}}\sqrt{(n+1)2e\hbar B},n=0,1,2,...
\label{eqC10}
\end{eqnarray}

Electrons have positive energies values, while holes have negative energies. Since the above relations are spin-independent,  the Landau surfaces form dual degeneracy. Furthermore, the energy levels are not evenly spaced in real-space. This can be inferred by regarding the eigenfunctions of the system which are given by solving Eq. \eqref{eqC2} and are known as Landau levels: 
\begin{eqnarray}
{{\varphi }_{{{k}_{B}}}}\propto {{e}^{\frac{-eB{{(y-{{y}_{0}})}^{2}}}{2\hbar }}}{{H}_{n}}[(y-{{y}_{0}})\sqrt{\frac{eB}{\hbar }}].   
\label{eqC11}           
\end{eqnarray} 
Here, one can see that the center of the Landau level in real space is given by $y_0$, inversely proportional with the magnetic field. Meanwhile, the radius of the Landau level is approximately given by $2\sqrt(\hbar/eB)$. This means that high effective magnetic fields concentrate the electronic population at the center of bump.

\bibliographystyle{apsrev4-2}
\bibliography{references}

\end{document}